\documentclass[usenatbib]{mn2e}

\pdfoutput=1

\usepackage{url}
\usepackage{fixltx2e}
\usepackage{mathptmx}
\usepackage[pdftex]{graphicx}
\usepackage{color}

\newcommand{\Mpc}{\rm\thinspace Mpc}

\newcommand{\km}{\rm\thinspace km}

\newcommand{\cm}{\rm\thinspace cm}
\newcommand{\yr}{\rm\thinspace yr}

\newcommand{\s}{\rm\thinspace s}

\newcommand{\Msun}{\hbox{$\rm\thinspace M_{\odot}$}}

\newcommand{\Msunpyr}{\hbox{$\Msun\yr^{-1}\,$}}

\newcommand{\keV}{\rm\thinspace keV}

\newcommand{\erg}{\rm\thinspace erg}

\newcommand{\ergpcmsqps}{\hbox{$\erg\cm^{-2}\s^{-1}\,$}}

\newcommand{\kmps}{\hbox{$\km\s^{-1}\,$}}

\newcommand{\kmpspMpc}{\hbox{$\kmps\Mpc^{-1}$}}

\newcommand{\Zsun}{\hbox{$\thinspace \mathrm{Z}_{\odot}$}}

\newcommand{\psqcm}{\hbox{$\cm^{-2}\,$}}

\begin{document}

\title[X-ray surface brightness fluctuations in AWM\,7]{Deep
  \emph{Chandra} and \emph{XMM-Newton} X-ray observations of AWM\,7 --
  I: Investigating X-ray surface brightness fluctuations}

\author
[J.~S. Sanders and A.~C. Fabian]
{J.~S. Sanders and A.~C. Fabian
  \\
  Institute of Astronomy, Madingley Road, Cambridge. CB3 0HA\\
}
\maketitle

\begin{abstract}
  We investigate the levels of small scale structure in surface
  brightness images of the core of the X-ray bright cool-core galaxy
  cluster AWM\,7. After subtraction of a model of the smooth cluster
  emission, we find a number of approximately radial surface
  brightness depressions which are not present in simulated images and
  are seen in both the \emph{Chandra} and \emph{XMM-Newton} data.  The
  depressions are most strongly seen in the south of the cluster and
  have a magnitude of around 4 per cent in surface brightness. We see
  these features in both an energy band sensitive to the density (0.6
  to 5~keV) and a band more sensitive to the pressure (3.5 to
  7.5~keV).  Histograms of surface brightness in the data, when
  compared to realisations of a smooth model, reveal stronger surface
  brightness variations. We use the $\Delta$-variance technique to
  characterise the magnitude of the fluctuations as a function of
  length scale. We find that the spectrum in the 0.6 to 5 keV band is
  flatter than expected for Kolmogorov index fluctuations. If
  characterised by a power spectrum, on large scales it would have an
  index around $-1.7$, rather than $-3.7$. The implied
  3D density fluctuations have a standard deviation of around 4 per
  cent. The implied 3D pressure variations are at most 4
    per cent.  Most of the longer-scale power in the density spectrum
  is contributed by the southern half of the cluster,
    where the depressions are seen. The density variations implied by
  the spectrum of the northern sector have a standard deviation of
  about 2 per cent.
\end{abstract}


\section{Introduction}
The velocity structure and degree of turbulence in the intracluster
medium (ICM) of galaxy clusters is a matter of much debate as there
are few direct observational constraints on its dynamical
properties. The nonthermal contributions to gas pressure, if
significant, are important to include for the determination of galaxy
cluster masses from X-ray data.  The viscosity and turbulence of the
ICM have a bearing on how energy can be dissipated within the ICM,
which connects to how the AGN can feed energy back into its
environment \citep[e.g.][]{Kunz11}.

Theoretical calculations predict that the fraction of pressure support
in the ICM from gas motions, generated by cluster growth and mergers,
ranges from 5 to 20 per cent in the centres, increasing at larger
radii (e.g. \citealt{Lau09, Vazza09, Vazza11}). Relaxed clusters have
significantly lower amounts of turbulence than those merging or have
recently merged, with turbulent pressure fractions of less than 10 per
cent. However, the amount of turbulence in simulations depends
crucially on transport processes such as viscosity or
magnetohydrodynamical effects. The active galactic nuclei (AGN) in the
centres of clusters are also often seen to be affecting the
surrounding ICM, injecting jets and bubbles into the cluster
\citep[e.g.][]{McNamaraNulsen07} and likely causing gas motions. These
generated motions have been predicted to range from 500 to $1000\kmps$
\citep{Bruggen05,Heinz10}.

One direct method of measuring or limiting gas motions in galaxy
clusters is to examine the width of X-ray emission lines using the
X-ray reflection grating spectrometer (RGS) on \emph{XMM-Newton}. We
recently placed a limit on ICM random gas motions in the core of Abell
1835 of $274 \kmps$, at the 90 per cent level
\citep{Sanders10_A1835}. The implied limit on the turbulent to thermal
energy density is less than 13 per cent. For a sample of objects we
were able to find at least 15 sources with less than 20 per cent of
the thermal energy density in turbulence
\citep{Sanders10_Broaden}. This technique is, however, limited to
sources with compact X-ray emitting cores and we can only examine the
innermost regions. Other techniques to obtain limits or measure the
velocity structure in clusters or elliptical galaxies includes
examining the degree of resonance scattering in high resolution X-ray
spectra \citep{Werner09}, which imply strong upper limits on the gas
motions.

Each of the above probes for gas motions indicates rather small
amounts of gas motion in the cluster cores. There is additional
indirect evidence that gas motions in the ICM are not strong in
relaxed objects. These include the smooth changes in velocity of the
filaments in the core of the Perseus cluster \citep{HatchPer06}, which
are apparently tracing gas motion \citep{FabianPerFilament03} and
their linear morphology.

Comparisons of X-ray and optical gravitational profiles in elliptical
galaxies indicate additional non-thermal pressure contributions, some
of which may be due to gas motions. The non-thermal pressure is around
10 per cent of the thermal gas pressure in M\,87 and NGC\,1399
\citep{Churazov08} or 20 to 30 per cent in a sample
\citep{Churazov10}. In the Perseus cluster, \cite{Churazov04} found
evidence from the existence of resonance line emissions for gas
motions at around half the sound speed.

\cite{Schuecker04} noted that for isotropic turbulence, like the
velocity distribution, the pressure energy spectrum should have a
powerlaw dependence with wave number. For the pressure, $E_P(k)
\propto k^{-7/3}$ \citep[see for example, section 6.9
in][]{LesieurTurbFluids08}, rather than the standard 5/3 Kolmogorov
relation for velocity. In three-dimensions (3D) this corresponds to a
pressure power spectrum with a spectral index of
$-13/3$. \cite{Schuecker04} examined the spectrum of the fluctuations
in a pressure map of the Coma cluster. They found that the spectrum
between length scales of 40 and 90 kpc was well described by a
projected Kolmogorov type spectrum. Integrating the spectrum after
deprojection implied a lower limit of 10 per cent of the total ICM
pressure in turbulence.

\cite{ChurazovComa11} recently reexamined \emph{Chandra} and
\emph{XMM-Newton} data of the Coma cluster, finding that density
variations in this object are rather small. The amplitude of the
density fluctuations varies from 7 to 10 per cent on 500~kpc scales
down to 5 per cent on scales of around 30~kpc.

In the interstellar medium, the electron density power spectrum as
measured with interstellar scintillation is close to the value for a
Kolmogorov process ($-11/3$) over at least five orders of magnitude
\citep[e.g.][]{Armstrong95}. The spectral slope when observing
H\textsc{i} 21cm line emission can vary, however. The value for the
Small Magellanic Cloud is $-3.04$ \citep{Stanimirovic99}, but
measurements of the interstellar medium of dwarf galaxies gives slopes
from $-1.5$ to $-2.6$ \citep{Dutta09}.

Simulations of density fluctuations using hydrodynamics \citep{Kim05}
found a Kolmogorov slope for low Mach numbers, flattening as the Mach
number increased. For 3D compressible magnetohydrodynamical
simulations of turbulence \cite{Kowal07} found that the energy
spectrum of density fluctuations with subsonic turbulence is
consistent with $-7/3$ for strong magnetic fields and the Kolmogorov
index of $-5/3$ for weak magnetic fields. For supersonic turbulence
the density spectrum flattens, like the hydrodynamic simulations.

In this paper we use deep X-ray observations of a nearby ($z=0.0172$)
galaxy cluster AWM\,7 to look for fluctuations in surface brightness
which can indicate the degree of gas motions or turbulence. AWM\,7 is
an X-ray bright ($F_{2-10\keV} \sim 9 \times 10^{-11} \ergpcmsqps$;
\citealt{Edge92}) poor cluster of galaxies located close to the edge
of the Perseus galaxy cluster. The X-ray surface brightness is
elongated along the east-west direction in the same direction as the
Perseus-Pisces chain of galaxies \citep{Neumann95}. This is claimed to
be caused by the infall of material along that direction rather than
tidal effects. There is evidence for a clump of galaxies to the east
of the cluster \citep{Beers95}. The central dominant galaxy and
central X-ray peak is offset 30~kpc to the west of the centre of the
larger scale emission and the galaxy is aligned close to
perpendicularly to the X-ray emission \citep{Neumann95}. The redshift
of the central galaxy agrees with the cluster redshift and the cluster
has a velocity dispersion of $\sim 730\kmps$ \citep{Koranyi00}. No
significant velocity substructure is found \citep{Koranyi02}.

\cite{Sato08} found no evidence for bulk motions greater than the
sound speed $\sim 1000\kmps$, using the \emph{Suzaku}
observatory. They found that the ICM temperature declines from a peak
of 3.85~keV to 3.35~keV at 500~kpc. The cluster has a steeply peaked
surface brightness profile, from which would be inferred a cooling
rate of 20--35$\Msunpyr$ in the absence of heating
\citep{Peres98,Neumann95}, and a cool 2 keV core
\citep{Furusho03}. The central mean radiative cooling time is short at
$0.4$~Gyr \citep{Cavagnolo09}.

In this paper we restrict ourselves to an analysis of the surface
brightness of the galaxy cluster AWM\,7, leaving a spectral analysis
to a future work.  We assume, unless stated otherwise, that $H_0 =
70\kmpspMpc$. At a redshift of 0.0172, 1 arcsec on the sky corresponds
to 0.351~kpc.

\section{Observations}
\subsection{Processing datasets}
\label{sect:obs}
\begin{table}
  \centering
  \begin{tabular}{llll}
    \hline
    Observatory       & Observation& Date       & Exposure (ks) \\ \hline
    \emph{Chandra}    & 11717      & 2009-11-01 & 39.6 \\
    \emph{Chandra}    & 12016      & 2009-11-03 & 22.3 \\
    \emph{Chandra}    & 12017      & 2009-11-05 & 45.1 \\
    \emph{Chandra}    & 12018      & 2009-11-06 & 26.8 \\
    \emph{XMM-Newton} & 0135950301 & 2003-02-02 & 31.6 \\
    \emph{XMM-Newton} & 0605540101 & 2009-08-16 &  126 \\
    \hline
  \end{tabular}
  \caption{X-ray observations examined in this paper. Shown are the observatory,
    observation identifier, date of observation and exposure time. The
    exposure time shown is the full exposure for \emph{Chandra}
    as all the data were used. For \emph{XMM} the exposure is the
    average cleaned MOS exposure time.}
  \label{table:observations}
\end{table}

The galaxy cluster AWM\,7 has been observed by both \emph{Chandra} and
\emph{XMM-Newton}. Table \ref{table:observations} shows the
observations we analyse in this paper. The \emph{Chandra} observations
were taken using the ACIS-I detector. All of the \emph{Chandra}
observations use the same aim-point. The \emph{XMM-Newton}
observations included the EPIC-MOS and PN detectors and the RGS
instruments. We will mostly examine the \emph{Chandra} data of the
object in this paper, but also include the \emph{XMM} EPIC-MOS data
for confirmation. Although the EPIC-PN data could be used, the
detector has numerous large chip gaps and a higher background than the
MOS detectors, which make it more difficult to interpret.

We processed each set of data using standard software packages. The
\emph{Chandra} data were processed using the \textsc{ciao} tool
\textsc{acis\_process\_events}, using very faint mode grade rejection
to reduce the background.  The entire field of view of the ACIS-I
detector was filled by the cluster and no additional front-illuminated
CCDs were enabled. We therefore extracted lightcurves to check for
flares from the edge of the ACIS-I array, as far as possible from the
core of the cluster. We examined lightcurves in different energy
filters (0.5--12, 0.5--2 and 2--12~keV) by eye, finding no evidence
for any flaring. The different observations had consistent count
rates. We therefore did not filter the \emph{Chandra} datasets for
flaring. Each event file was reprojected so that the physical
coordinates matched the 12017 observation.

The \emph{XMM-Newton} data were processed with \textsc{sas} pipeline
\textsc{emchain}. We filtered flares in the data using the lightcurve
in the 10 to 15 keV band for events with a pattern of zero. Periods
above 0.35 count per second were excluded for the MOS detectors. The
coordinate system for dataset 0135950301 was reprojected to match the
observation 0605540101. We used events with pattern values of 12 or
less for the MOS detectors.

Point sources in the data were excluded from the data by eye. We
examined the 0.5--2, 2--7 and 0.5--7~keV bands for point sources. We
applied smoothing to identify fainter point sources, using smoothing
scales of 6 and 8~arcsec. The excluded sources are shown as grey
circles in Figure~\ref{fig:xmm_chandra_compar}.

\subsection{Background data}
\begin{table}
  \centering
  \begin{tabular}{ll}
    \hline
    Observatory & Observations \\
    \hline
    \emph{Chandra} & acis0iD2005-09-01bkgrnd\_ctiN0001.fits \\
    & acis1iD2005-09-01bkgrnd\_ctiN0001A.fits \\
    & acis2iD2005-09-01bkgrnd\_ctiN0001.fits \\
    & acis3iD2005-09-01bkgrnd\_ctiN0001.fits \\
    \hline
    \emph{XMM-Newton} 
    & 1468\_0504102401 MOS1 (-4), MOS2 (-5) \\
    & 1500\_0511011001 MOS1, MOS2 (-5) \\
    & 1512\_0505480701 MOS1 (-4), MOS2 (-5) \\
    & 1548\_0556200201 MOS1, MOS2 \\
    & 1569\_0553912001 MOS1 (-4), MOS2 \\
    & 1587\_0556214601 MOS1, MOS2 \\
    & 1595\_0551270201 MOS1 \\
    & 1607\_0510780201 MOS1 (-5), MOS2 (-5) \\
    & 1796\_0510780401 MOS1 (-5), MOS2 (-5) \\
    & 1844\_0604890201 MOS1 (-4), MOS2 \\
    & 1935\_0653870701 MOS1 (-4), MOS2 (-5) \\
    & 2027\_0412580701 MOS1, MOS2 \\
    \hline
  \end{tabular}
  \caption{Background datasets used in this analysis. The
    \emph{Chandra} data are standard blank-sky backgrounds identified
    using \textsc{acis\_bkgrnd\_lookup}. The \emph{XMM-Newton}
    observations are closed-shutter particle background observations taken from
    \protect\url{http://xmm.vilspa.esa.es/external/xmm_sw_cal/background/filter_closed/}.
    Shown are which MOS detector datasets were taken from each
    observation. In brackets are listed the excluded anomalous-state CCDs.}
  \label{tab:background}
\end{table}

In order to get a good measurement of the surface brightness
background subtraction can be important. For the \emph{Chandra} data
we used blank-sky background files (listed in Table
\ref{tab:background}). We combined the four background event files for
the individual CCDs, reprojecting them to match the aspect of the
observations. By adjusting the background exposure time, we normalised
the count rate in the 10 to 12 keV band to match the observations.

The standard background files subtracted too much from the data
however, leaving clear residuals. This was most apparent in the
source-free 9 to 12 keV band. We found that the residuals were due to
a mismatch between the bad pixels in the background and observation
datasets. The bad pixel maps of the observations contained several
more rows of bad pixels than were excluded from the standard
backgrounds. To fix this, we created a list of the bad detector pixels
which were common to the AWM~7 observations. We rejected any events in
the background event files which had detector pixel coordinates in
this list. After this filtering, in the high energy band there was a
close match between the data and background files as expected.

The background subtraction of the \emph{Chandra} data is not a
critical part of the analysis, however. If no background subtraction
is done, the features seen in surface brightness images do not
change. In addition the spectra of fluctuation signal as a function of
length scale (Section \ref{sect:deltavar}) do not change
significantly.  The use of blank sky backgrounds is not strictly
correct, however, as AWM~7 lies in a region of low galactic
latitude. The soft part of the background spectrum, particularly after
the high energy count rate renormalisation, may be
incorrect. Fortunately AWM~7 is a rather bright object. The 3.5 to
7.5~keV we examine later is the most sensitive to the background, but
in this case it is dominated by particle background which we correctly
subtract.

For the \emph{XMM-Newton} data, which we examine only qualitatively,
we subtracted the particle part of the background. We took recent
closed-shutter EPIC-MOS observations (Table~\ref{tab:background}),
excluding CCDs which were in the anomalous state
\citep{KuntzSnowden08}. We filtered these data in the same way as the
main observations and reprojected them to match the 0605540101
observation. A copy of this background was then reprojected to match
the 0135950301 observation. We normalised the exposure of the
background data in each CCD to match each observation count rate in
the 9 to 12 keV band. We combined images of these background files in
the same proportion as the 0135950301 and 0605540101 exposure
times. We concentrate on the central MOS CCDs which are not affected
by the anomalous state.

\section{Images}
\begin{figure}
  \centering
  \includegraphics[width=\columnwidth]{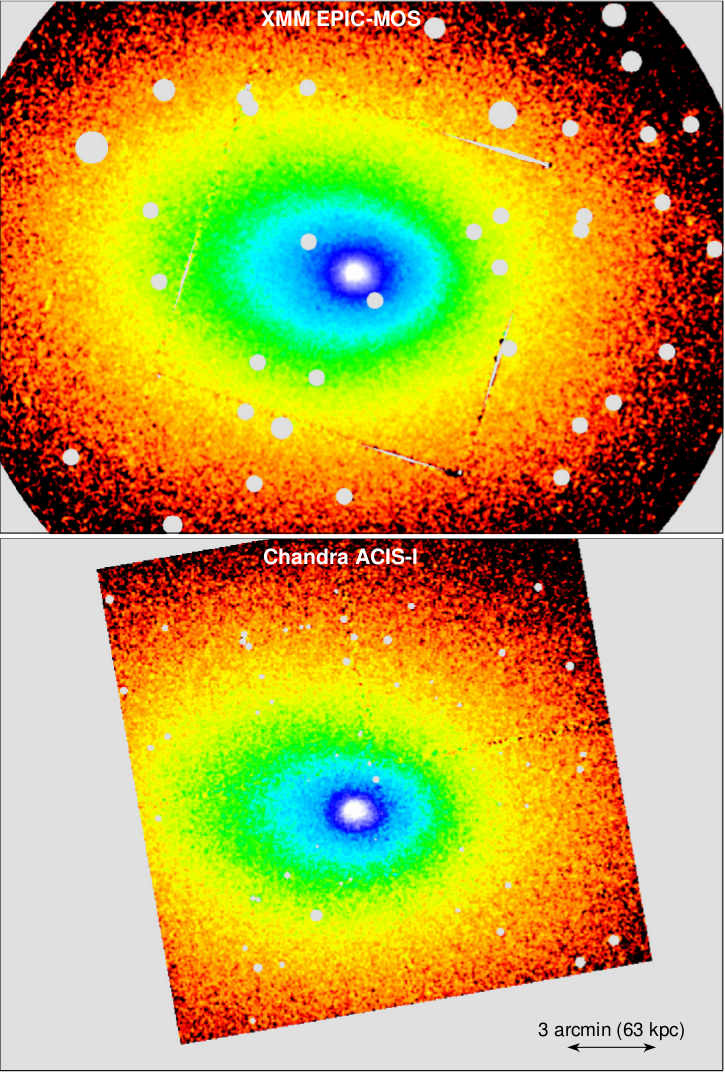}
  \caption{\emph{XMM} EPIC-MOS and \emph{Chandra} 
      background-subtracted and exposure-corrected images of the
    cluster in the 0.6 to 5 keV band. The \emph{XMM} data have a
    binning of 1.25~arcsec per pixel and the \emph{Chandra}
    1.968~arcsec. The data were smoothed by Gaussians of
      $\sigma=2$ and 1 pixels, for \emph{XMM} and \emph{Chandra},
    respectively. Excluded point sources and regions are shown in
    grey.}
  \label{fig:xmm_chandra_compar}
\end{figure}

Fig.~\ref{fig:xmm_chandra_compar} shows the combined
background-subtracted and exposure-corrected images of the cluster
using the \emph{XMM} EPIC-MOS (top panel) and \emph{Chandra} data
(bottom panel) in the 0.6 to 5 keV band.  There are $6.9\times10^5$
events in this energy range in the central 3.5 arcmin radius in the
\emph{Chandra} data and $1.3\times10^6$ in \emph{XMM} EPIC-MOS. For
the \emph{Chandra} data we exposure-correct using exposure maps
weighted in energy using a 3.75~keV \textsc{apec} $0.5\Zsun$ spectral
model \citep{SmithApec01}, with an appropriate redshift and Galactic
absorption.  For \emph{XMM-Newton} we produce exposure maps in the
range of 0.8 to 1.6 keV around the Fe-L peak.

\begin{figure*}
  \centering
  \includegraphics[width=\columnwidth]{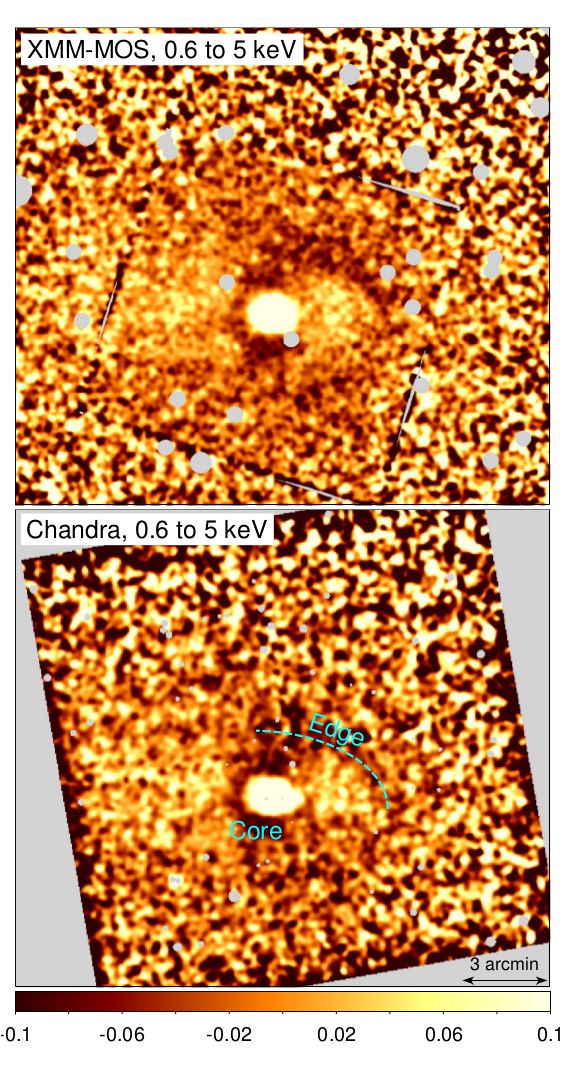}
  \hspace{3mm}
  \includegraphics[width=\columnwidth]{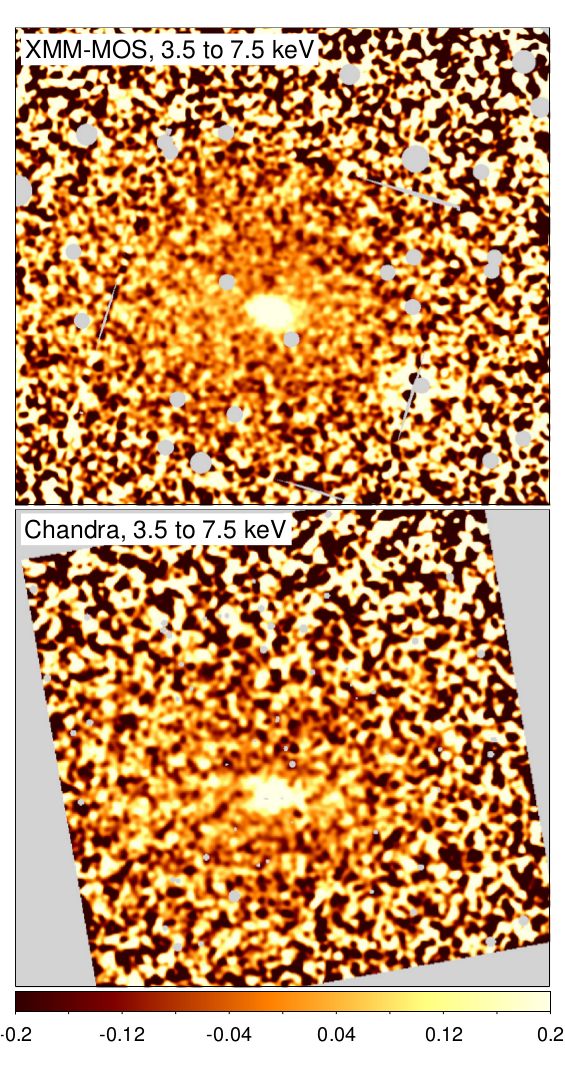}
  \caption{Unsharp-masked images of the cluster, in the
    density-sensitive 0.6--5~keV (left panels) and pressure-sensitive
    3.5--7.5~keV bands (right panels). The top panels show the
    fractional difference between \emph{XMM} EPIC-MOS data smoothed by
    Gaussians with $\sigma=4$ and 32 1.25 arcsec pixels.  The bottom
    panels show the results for \emph{Chandra}, showing the fractional
    difference between images smoothed by Gaussians of 3 and 20 1.968
    arcsec pixels. The bright central region is the central core which
    is saturated above the range shown.}
  \label{fig:unsharp}
\end{figure*}

As can be seen in these images, the cluster has a smooth surface
brightness profile. There is, however, structure within 1 arcmin
radius of the central nucleus and there is a jump in the surface
brightness profile around 3 arcmin to the west of the nucleus. This
western edge can be seen in unsharp-masked images of the cluster in
this band (Fig.~\ref{fig:unsharp} left panels; marked as Edge) and may
be a poorly defined cold front.

\begin{figure}
  \centering
  \includegraphics[width=\columnwidth]{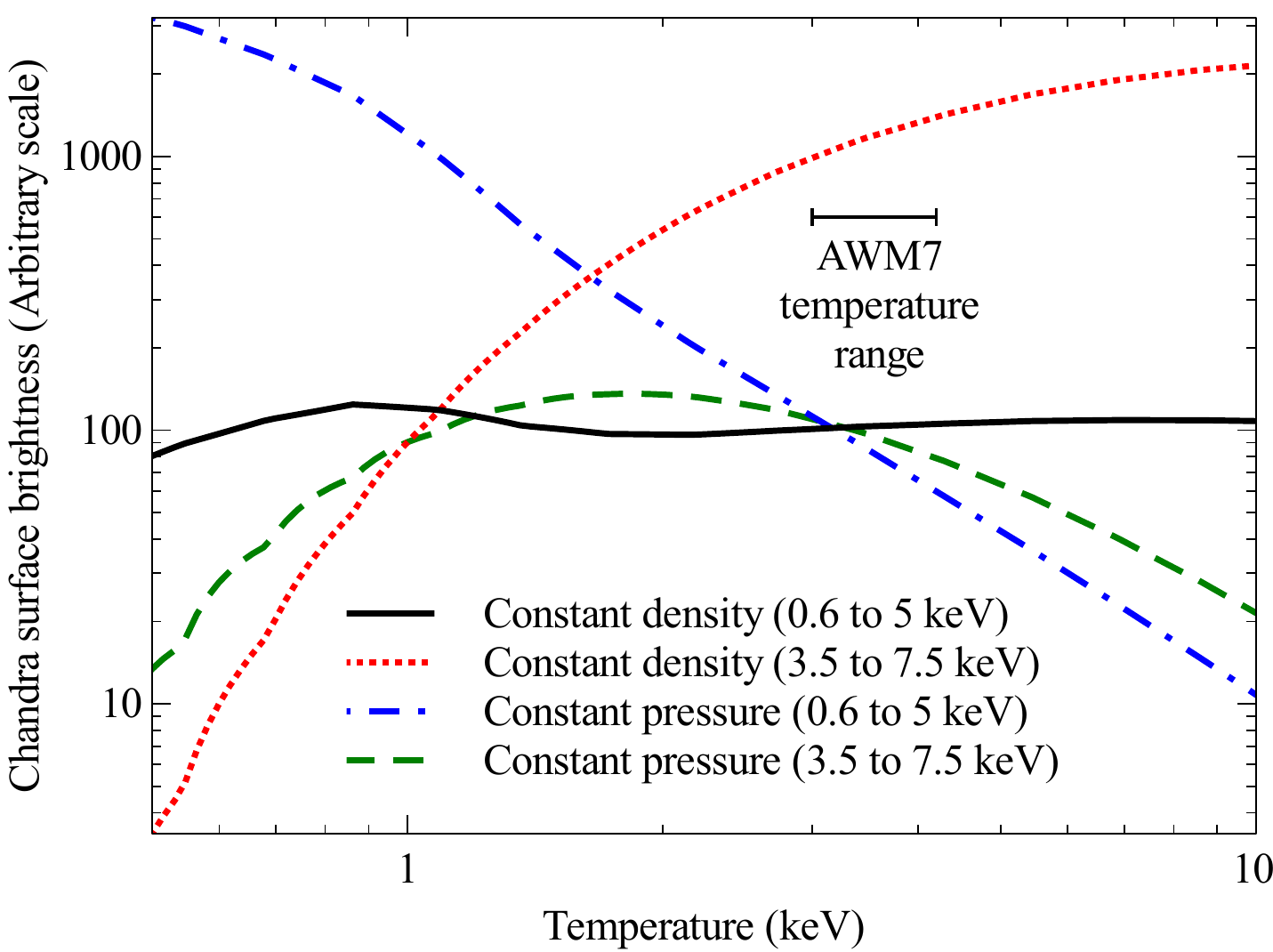}
  \caption{Variation in surface brightness as a function of
    temperature at constant density or pressure \protect\citep[similar
    to Figure 1 in][]{FormanM8707}. The surface brightness in two
    bands, 0.6 to 5~keV and 3.5 to 7~keV are shown. The lines have
    arbitrary normalization. We also show the range of projected
    temperature in the region examined by our observations outside of
    the central core.}
  \label{fig:ctdepend}
\end{figure}

To understand the effect of turbulence and gas motions, we would
ideally want to study the 3D distribution of properties of the ICM,
including the velocity, pressure, density, temperature and
metallicity. Unfortunately these quantities are difficult to
measure. The most accessible quantity is the X-ray surface
brightness. The bolometric surface brightness closely probes the
square of the density, integrated along the line of sight.

The energy band used is sensitive to different plasma properties,
however. The 3.5 to 7.5~keV energy band can be used to examine
variations in the integral of pressure-squared along the line of
sight for temperatures between 1 and 3 keV \citep{FormanM8707}. In
Fig.~\ref{fig:ctdepend} we show the temperature dependence of the
surface brightness in this hard band and in a wider 0.6 to 5~keV band,
at constant density and at constant pressure. For this simulation we
used a Galactic column density of $10^{21} \psqcm$ and the redshift of
AWM\,7. Typical temperatures in AWM\,7 are $\sim 3.75$~keV, so the 3.5
to 7.5~keV energy band is not purely sensitive to the pressure-squared
and has some temperature dependence. It is a better pressure-squared
proxy than the 0.6 to 5 keV band. The 0.6 to 5~keV band has very
little temperature dependence at constant density and is a good proxy
for the integral of the density-squared along the line of sight.

We show in Fig.~\ref{fig:unsharp} (right panels) an unsharp-masked
image in the 3.5 to 7.5~keV band, with the same smoothing parameters
as for the 0.6 to 5~keV images in the left panels.  The signal to
noise ratio is lower in this image and we do not clearly see the edge
that is apparent in the 0.6 to 5~keV unsharp-masked image. This may be
because this band is more pressure sensitive and features are contact
discontinuities, or that the signal to noise ratio is lower.

\section{Fluctuations in surface brightness}
We would like to remove the large scale cluster structure to examine
the surface brightness variation on smaller scales.  As the cluster is
rather smooth and free of substructure, we model it by fitting
ellipses to contours of surface brightness, spaced
logarithmically. Model values for each pixel are calculated by
linearly interpolating in logarithmic space the surface brightness
between the two neighbouring ellipses. We mask out point sources and
excluded sky regions in both the real and model images.  We call the
ellipse fitting and interpolation model ELFIT for the purposes of this
paper. The ELFIT model removes both azimuthally symmetric emission and
edge features, such as cold fronts, making it ideal to model the more
subtle differences due to turbulent motions. In the 0.6 to 5 keV band
we fit 20 ellipses and 10 in the 3.5 to 7.5 keV band.

\begin{figure}
  \centering
  \includegraphics[width=\columnwidth]{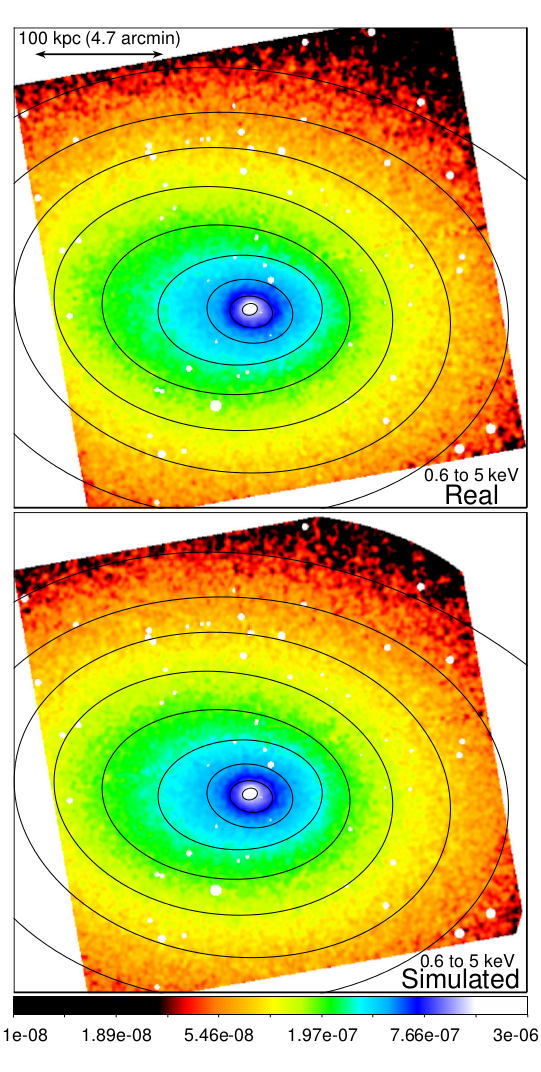}
  \caption{Real and simulated background-subtracted,
    exposure-corrected images of the cluster in the 0.6 to 5~keV
    band. The simulated data are a realisation of the ellipse fitting
    ELFIT model. The images have 1.968 arcsec binning and are smoothed
    by a Gaussian of $\sigma=2$ pixels. The ellipses shown are fitted
    to the surface brightness and listed in Table \ref{tab:ellipses}.}
  \label{fig:sim_vs_real}
\end{figure}

To construct a simulated image of the cluster, we take the smooth
ELFIT model of the cluster, multiply by the exposure map of the real
observation, add a background component and then make a Poisson
realisation. This model dataset can be compared to the observed
surface-brightness by background-subtracting and exposure-correcting.
We can include fluctuations in the ELFIT model to see how detectable
they are. Fig.~\ref{fig:sim_vs_real} shows the real
background-subtracted, exposure-corrected 0.6 to 5~keV and a simulated
image of the cluster without any additional fluctuations. The ELFIT
model does a good job at reproducing the overall surface brightness
distribution.  In our images we use a binning of four \emph{Chandra}
detector pixels (1.968 arcsec) per image pixel.

\begin{table}
  \centering
  \begin{tabular}{llllll}
    \hline
    RA & Dec & Radius & Radius   & Angle & $<r>$ \\
    \hline
    43.614 & 41.579 & 13  & 17   & 103.3 & 3.7 \\
    43.613 & 41.577 & 35  & 47   & 81.2  & 10.9 \\
    43.614 & 41.578 & 70  & 95   & 81.2  & 22.6 \\
    43.622 & 41.578 & 121 & 181  & 91.8  & 43.4 \\
    43.633 & 41.578 & 186 & 274  & 84.8  & 71.2 \\
    43.635 & 41.579 & 267 & 376  & 82.1  & 99.3 \\
    43.628 & 41.578 & 356 & 484  & 81.8  & 129 \\
    43.629 & 41.575 & 447 & 612  & 86.5  & 162 \\
    43.577 & 41.431 & 951 & 1144 & 40.6  & 189 \\
    \hline
  \end{tabular}
  \caption{Centres (J2000), radii (arcsec) and angles (deg) of the
    ellipses used for examining surface brightness distributions.
    We also show the mean radius in kpc from the cluster
    centre of the pixels between the ellipse
    and the ellipse interior to it (if any).}
  \label{tab:ellipses}
\end{table}

In Section \ref{sect:histo} we will examine histograms of the surface
brightness of the cluster as a function of radius. To do this
we divide the cluster into regions using ellipses fitted to contours
in surface brightness. The contours are placed logarithmically in
surface brightness in the 0.5 to 7~keV band. They are shown in
Fig.~\ref{fig:sim_vs_real} and their properties are listed in
Table~\ref{tab:ellipses}.

\begin{figure*}
  \centering
  \includegraphics[width=\columnwidth]{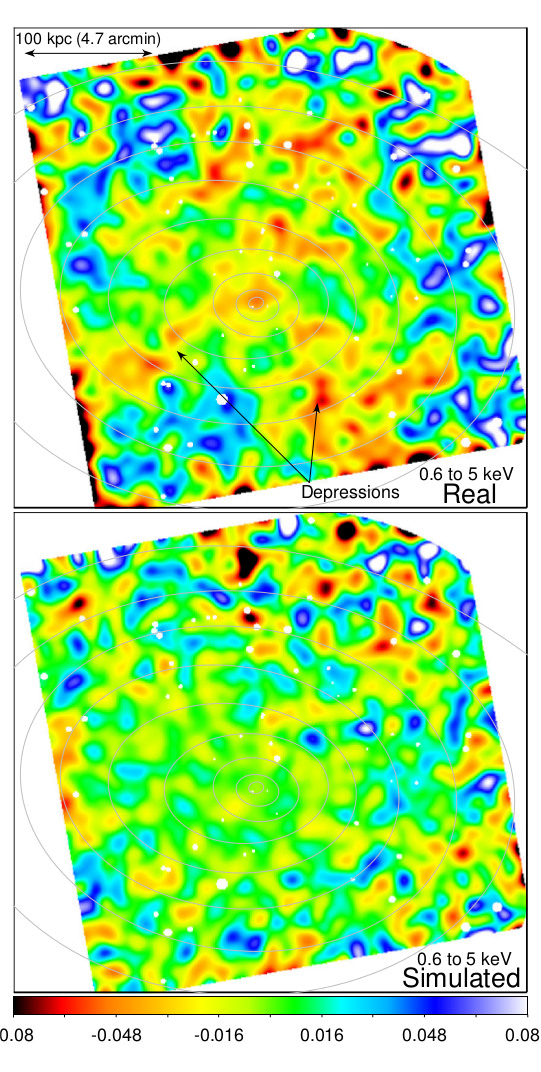}
  \hspace{3mm}
  \includegraphics[width=\columnwidth]{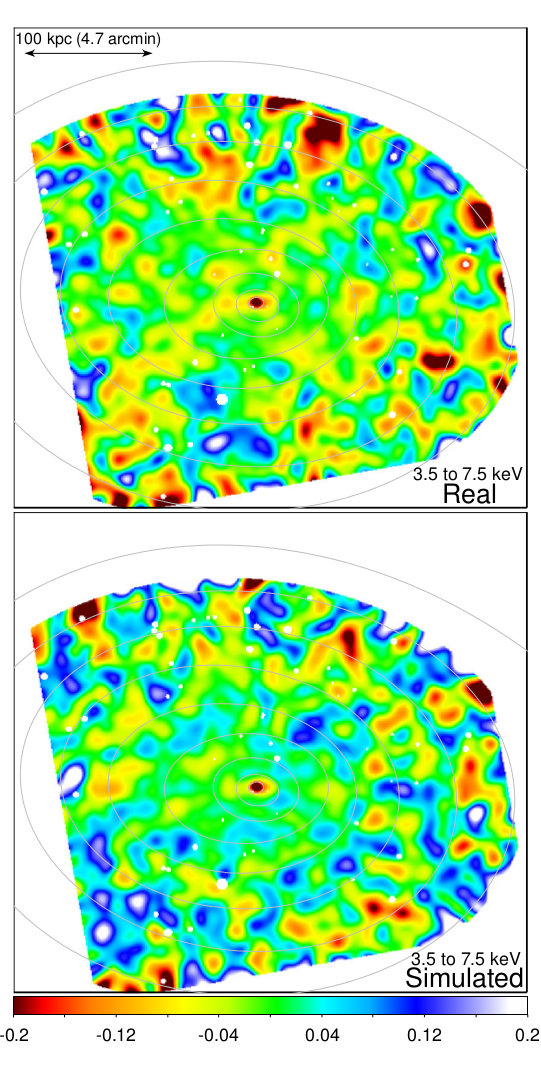}
  \caption{Fractional residuals between the
    \emph{Chandra} data and the smooth ELFIT model, after smoothing
    both images with a Gaussian of $\sigma=8$ pixels. The top panels
    show the results for the real data, whilst the bottom panels show
    the same analysis for a cluster simulated from the model. The left
    two panels show the residuals for the density-sensitive 0.6 to
    5~keV band and the right two panels show the pressure-sensitive
    3.5 to 7.5~keV range.}
  \label{fig:sim_div_real}
\end{figure*}

\subsection{Fluctuation images}
Images of the model ELFIT surface brightness can be compared to the
real surface brightness in detail. We smooth both the real data and
the model with a Gaussian of $\sigma = 8$ pixels (15.7~arcsec), and
then compute the fractional difference between the smoothed data and
smoothed model. We also make a simulated dataset using the ELFIT model
and compute the fractional difference between this simulation and the
model. Fig.~\ref{fig:sim_div_real} shows the fractional differences in
the 0.6 to 5 keV band of the real data (top-left panel) and simulated
data to the model (bottom-left panel). There are features in the real
data which are not visible in the simulated dataset. The most easily
seen features are the apparent radial depressions from the south east
of the image to the centre, and from the south-south-east to the
centre.

The 3.5 to 7.5~keV band is sensitive to the pressure-squared.
Fig.~\ref{fig:sim_div_real} (top-right panel) shows the fractional
residuals between the real data and its ELFIT model, smoothing both by
a Gaussian of the same size as was used for 0.6 to 5~keV. We also plot
the fractional differences between a realisation of this model and
itself in the lower-right panel.  These data are much noisier than the
0.6 to 5~keV band, due to the lower number of counts and the higher
level of background. Many of the features appear to match the wider
band data, however, in particular the troughs in surface brightness in
the south. The strength of the variations also appears to match the
0.6 to 5 keV band and is larger than the variations in the simulated
data. It should be noted, however, that the two bands are not
completely independent, although the wider band is dominated by softer
emission.

\begin{figure}
  \centering
  \includegraphics[width=\columnwidth]{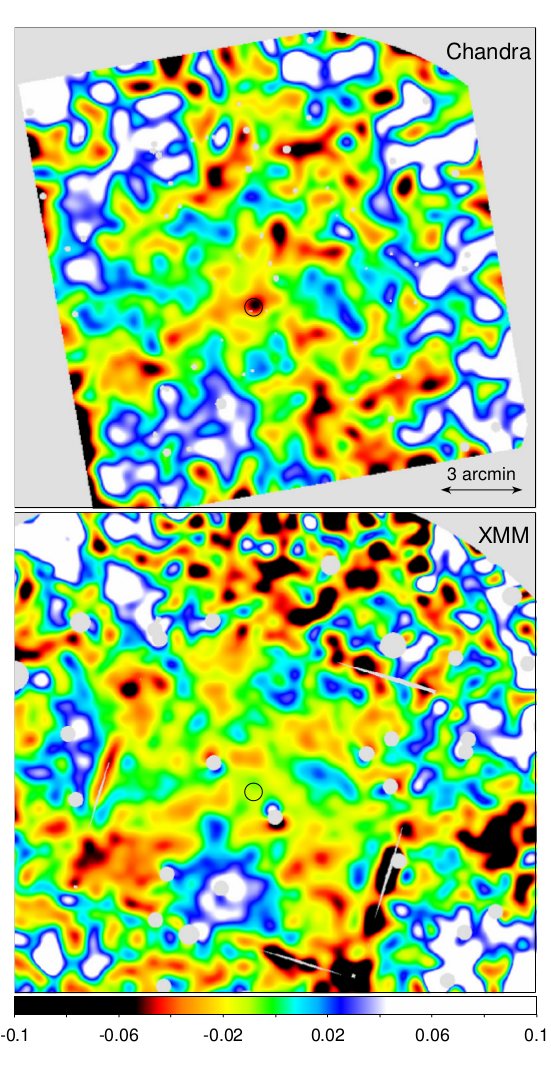}
  \caption{Fractional differences of the \emph{Chandra} and \emph{XMM}
    EPIC-MOS background-subtracted, exposure-corrected
    images to their respective ELFIT models in the 0.6 to 5 keV
    band. The data and models have been smoothed by 8 and 12 pixels,
    for \emph{Chandra} and \emph{XMM}, respectively. The black circle
    marks the X-ray peak of the cluster.}
  \label{fig:xmm_chandra_residuals}
\end{figure}

We can also examine the \emph{XMM-Newton} data to check that the
deviations are not instrument-dependent. In
Fig.~\ref{fig:xmm_chandra_compar} we showed the \emph{XMM} EPIC-MOS
combined background-subtracted, exposure-corrected image
of the cluster. We similarly model the MOS image of the cluster with a
20 ellipse ELFIT model. Fig.~\ref{fig:xmm_chandra_residuals} shows the
fractional residuals of the data to the model surface brightness,
after smoothing both by a Gaussian of 15 arcsec (12 pixels). It can be
seen that the \emph{Chandra} and \emph{XMM} residuals match each other
well, showing that many of the features are real deviations. The
magnitude of the depressions to the south is around 4 per cent.

\subsection{Pixel distribution histograms}
\label{sect:histo}

\begin{figure}
   \includegraphics[width=\columnwidth]{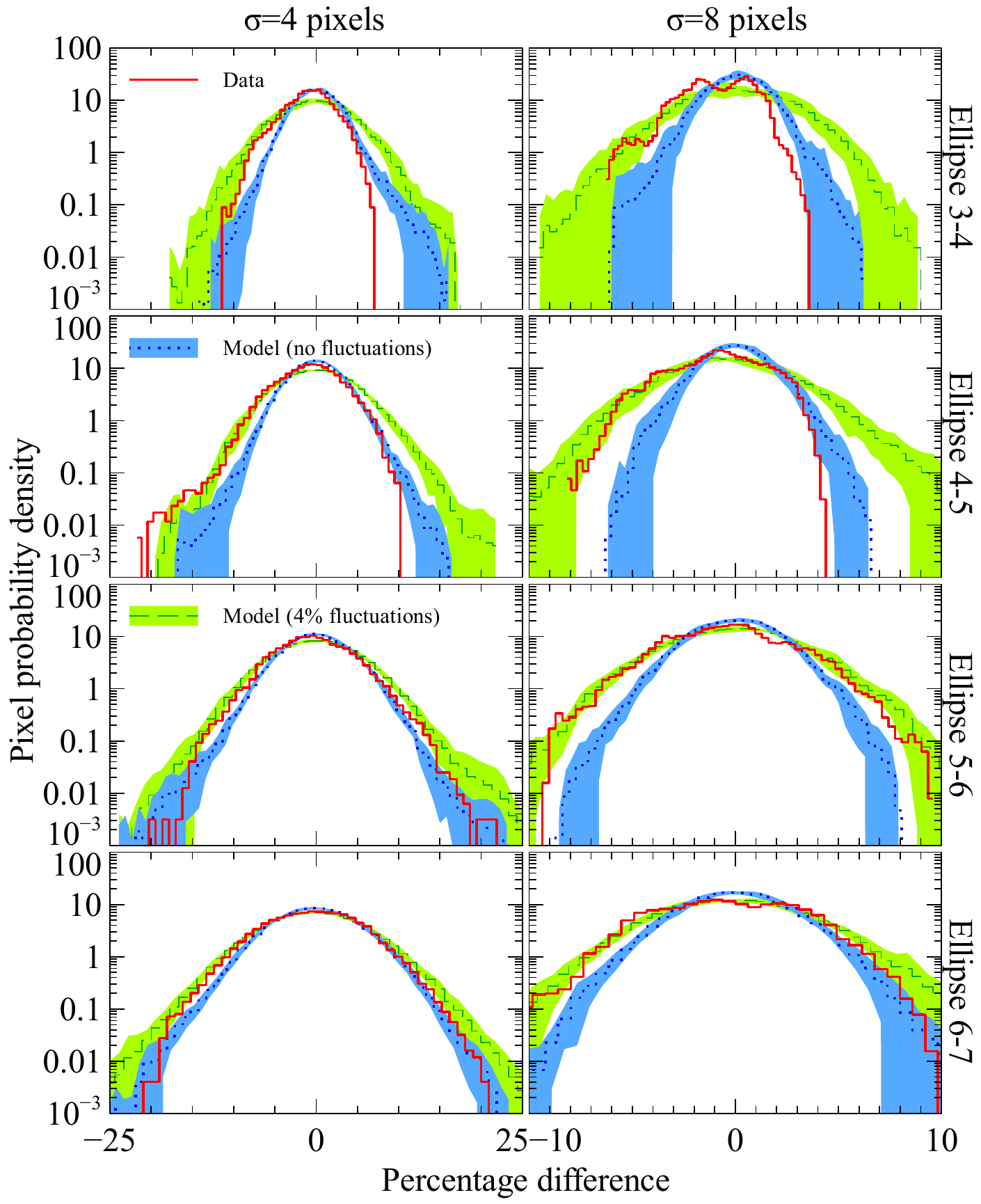}
   \caption{Distribution of surface brightness in the 0.6 to 5 keV
     band in different regions relative to the ELFIT model, after
     smoothing. The rows show the results between four pairs of
     ellipses, with increasing radius.  The columns show the results
     after smoothing with Gaussians with $\sigma=4$ and 8 pixels (2.8
     and 5.5 kpc). In each panel, the thick solid red line shows the
     real data. The dotted line and its surrounding blue shaded region
     shows the mean and standard deviation for realisations of the
     smooth ELFIT model itself.  The dashed line and its green shaded
     area shows the distribution for realisations of the ELFIT model
     plus Gaussian fluctuations of 4 pixel scale and a level of 4 per
     cent.}
   \label{fig:elldistn0650}
\end{figure}

A more quantitative measure of the structure in the surface brightness
images can be made by examining the distribution of surface brightness
as a function of radius. We investigate the distribution inside the
ellipses listed in Table~\ref{tab:ellipses} (shown in
Fig.~\ref{fig:sim_vs_real}).  We smooth the data and ELFIT model with
a Gaussian and compute a histogram of the fractional difference inside
four neighbouring pairs of the ellipses. Point sources and regions
outside that examined shown were filled with zero values before the
smoothing. We investigated two different Gaussian sizes, $\sigma=2.8$
and 5.6~kpc, sensitive to fluctuations on these scales.

The fractional difference histograms relative to the ELFIT model are
shown in Fig.~\ref{fig:elldistn0650}. The rows show the results in
ellipses increasing in radius and the columns show the effect of
smoothing with two different Gaussian widths.  The solid red line in
each plot shows the surface brightness deviations of the real data
relative to the ELFIT model. The dotted line and its blue surrounding
shaded region shows the mean and standard deviation of 10 realisations
of the ELFIT model to itself. The real data appear to have
substantially wider distributions of pixel values than these model
results. This difference is more apparent as more smoothing is
applied.

The dashed line and green shaded region shows the mean and standard
deviation of 10 realisations of the ELFIT model with Gaussian
fluctuations added, relative to the ELFIT model.  We generated
the Gaussian fluctuations by convolving an image where the pixels were
normally distributed, by a Gaussian of the size given, and scaling the
size of the resulting fluctuations to have the correct standard
deviation.  The magnitude of these fluctuations is chosen to have a
standard deviation of the model pixel values of 4 per cent. These
distributions better reproduce the width of the data distributions,
but the match is not exact.  If the surface brightness fluctuations
are of the order of 4 per cent, these correspond to 2 per cent
fluctuations in projected density.

\begin{figure}
  \includegraphics[width=\columnwidth]{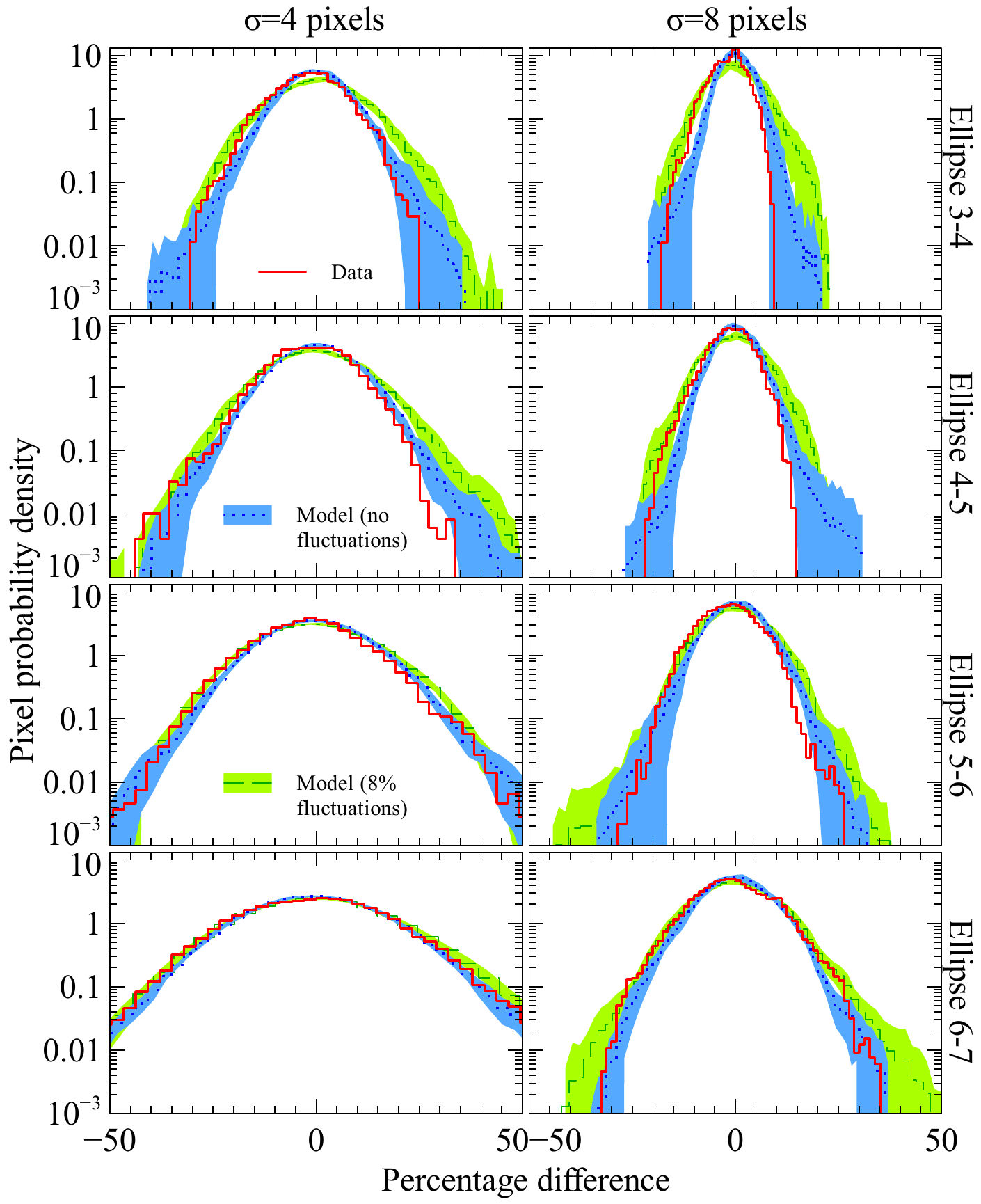}
  \caption{Distribution of surface brightness in the 3.5 to 7.5 keV
    band in different regions relative to the ELFIT model, plotted in
    the same way as Fig.~\ref{fig:elldistn0650}. The dashed line and
    shaded regions in this plot show a model with Gaussian
    fluctuations of 4 pixel scale and a level of 8 per cent added.}
  \label{fig:elldistn3575}
\end{figure}

We show the distributions for the 3.5 to 7.5~keV band in
Fig.~\ref{fig:elldistn3575}. The differences between the smooth model
and data are less apparent in this band, except when going to larger
radii and applying more smoothing. In this plot we also display the
results for a model with Gaussian fluctuations at the 8 per cent
level, instead of 4 per cent.

The additional width of the distributions compared to the model
confirm the fact that the maps appear to have additional fluctuations
in the image above that expected from Poisson noise.

\subsection{Cluster surface brightness modelling}
\begin{figure}
  \centering
  \includegraphics[width=0.85\columnwidth]{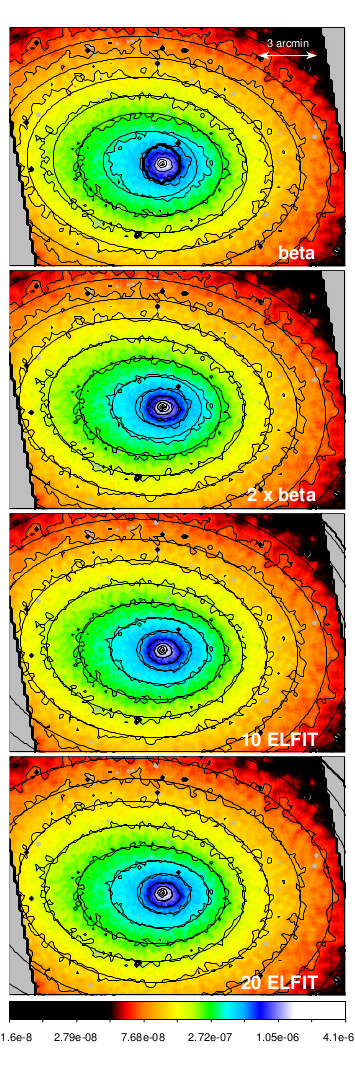}
  \caption{Adaptively-smoothed exposure-corrected \emph{Chandra} image
    in the 0.6 to 5 keV band. Contours of surface brightness are
    plotted (20 logarithmic contours from $1.2\times10^{-8}$ to
    $3\times10^{-6}$, appearing ragged). The smooth contours are of a
    fit using a $\beta$ model excluding the core (first row), a
    double-$\beta$ model including the core (second row), and 10 and
    20 ellipse ELFIT models (third and fourth rows, respectively).}
  \label{fig:contours}
\end{figure}

The fluctuations presented above were revealed by the subtraction of
an ELFIT cluster model. One potential problem with this approach is
that we could be subtracting real signal from the data by including it
in the model. The ELFIT model is good at removing features such as
edges, which typically follow surface brightness contours. The method
also removes shifts in the centres of isophotes as a function of
radius or isophotal twists.

In Fig.~\ref{fig:contours} is shown adaptively smoothed \emph{Chandra}
images of the cluster in the 0.6 to 5 keV band using the accumulative
smoothing algorithm described in \cite{SandersBin06}. These images
were produced using a variable-sized top-hat kernel with a radius
chosen to include at least 225 counts at each position. Plotted in
each image we display the jagged logarithmic-spaced surface brightness
contours.

\begin{figure}
  \centering
  \includegraphics[width=0.85\columnwidth]{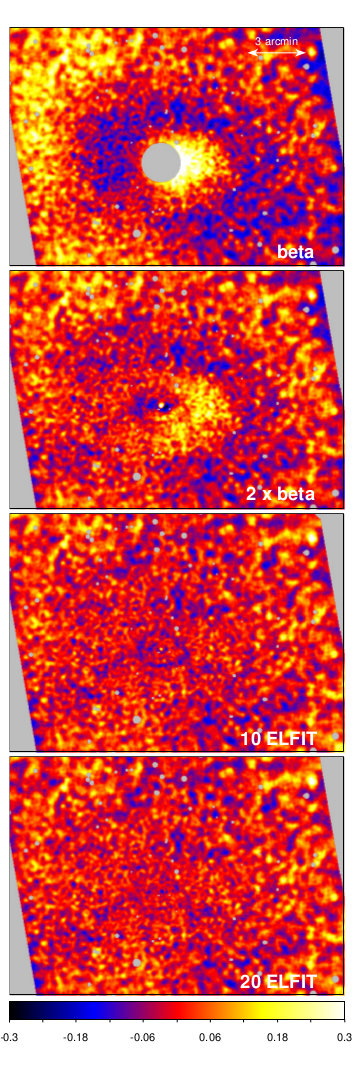}
  \caption{Fractional residuals between adaptively smoothed data and
    models, for the models shown in Fig.~\ref{fig:contours}.}
  \label{fig:modelratios}
\end{figure}

  In the top panel of Fig.~\ref{fig:contours}, we also show the
  contours of an elliptical $\beta$ model fitted to the surface
  brightness This model was fitted to the raw data using the
  \textsc{ciao} \textsc{sherpa} package, minimising the `cstat'
  statistic, which takes account of the Poisson distribution of the
  counts. We excluded the central arcminute of the cluster from the
  fit (shown as a bold circle). The $\beta$ model was allowed to have
  a variable ellipticity, centre, core radius and index. The
  functional form of the surface brightness model used was
  \begin{equation}
    S(x, y) = S(r) = S_{0}(1+[r/r_{0}]^{2})^{-\alpha}
  \end{equation}
  for power law index $\alpha$ and core radius $r_0$, using
  \begin{equation}
    r(x, y) = [x_2^2 (1-\epsilon)^2 + y_2^2]^{1/2} / (1-\epsilon),
  \end{equation}
  where $\epsilon$ is the ellipticity,
  \begin{equation}
    x_2 = (x-x_0)\: \mathrm{cos} \theta + (y-y_0)\: \mathrm{sin} \theta,
    \textrm{and}
  \end{equation}
  \begin{equation}
    y_2 = (y-y_0)\: \mathrm{cos} \theta - (x-x_0)\: \mathrm{sin} \theta,
  \end{equation}
  where $\theta$ is the angle of ellipticity and $(x_0, y_0)$ are the
  coordinates of the centre.

The model is a reasonable fit to the surface brightness on larger
scales (i.e. there is a close match between the contours
  of the data and model), but there are quite large offsets around
the core of the cluster between the model and data, particularly to
the west of the core and a depression beyond that.  These residuals
can be seen in the fractional difference between the adaptively
smoothed data and the model (Fig.~\ref{fig:modelratios}, top
panel). As noted by \cite{Neumann95}, the central X-ray peak is offset
from the outer parts of the cluster.

In the second panel of Fig.~\ref{fig:contours}, we show contours from
a two-component $\beta$ model fit to the surface brightness. In this
fit we did not exclude the central core. Each $\beta$ component was
allowed to have free centre, core radius, index and ellipticity during
the fitting.  The two-component model is a better fit to
  the data than the single-component model, although there are
discrepancies, particularly around 3 arcmin north of the core and 5
arcmin to the west. The residuals, shown in
Fig.~\ref{fig:modelratios}, are smaller than the single-component
$\beta$ model, but have a similar morphology.

The bottom two panels of Fig.~\ref{fig:contours} show the comparison
between ELFIT models using 10 and 20 ellipses. These
models are better fits than the $\beta$ models, as expected if the
isophotes are approximately elliptical. The residuals on larger
scales, shown in Fig.~\ref{fig:modelratios}, are substantially reduced
from the $\beta$ model fits. The 20 ellipse model is better at fitting
the edge an arcmin to the east of the core than the 10 ellipse model.

The residuals in Fig.~\ref{fig:modelratios} for the $\beta$ model fits
do not resemble the structures expected in the presence of simple
hydrodynamic turbulence (see Section \ref{sect:discuss}).

\subsection{Spectra of fluctuations}
\label{sect:deltavar}
It is useful to characterise the scale on which any fluctuations
occur.  The $\Delta$-variance method \citep{Stutzki98} measures the
variance in a two-dimensional image on a specific length
scale. The technique was recently used to examine X-ray
  observations of the Coma cluster \citep{ChurazovComa11}. In its
simplest form it works by filtering an image with a wavelet Mexican or
French hat function with a particular size and computing the variance
of the pixels. Pixels outside of the region of interest can be easily
excluded in the analysis by the use of a weighting map. The weighting
map is also convolved by the same function to allow the effect of any
edges to be removed.

We compare the $\Delta$-variance spectra with spectra of cluster
models, where we add different levels of fluctuations. We use two
different types of fluctuation field. The first fluctuation type is
simple Gaussian fluctuations on the projected surface
brightness. These are characterised by a Gaussian length scale,
$\sigma$, and the standard deviation of the fluctuation field,
expressed as a percentage on the underlying cluster surface brightness
(generated using the method described in Section \ref{sect:histo}).

\begin{figure}
  \includegraphics[width=\columnwidth]{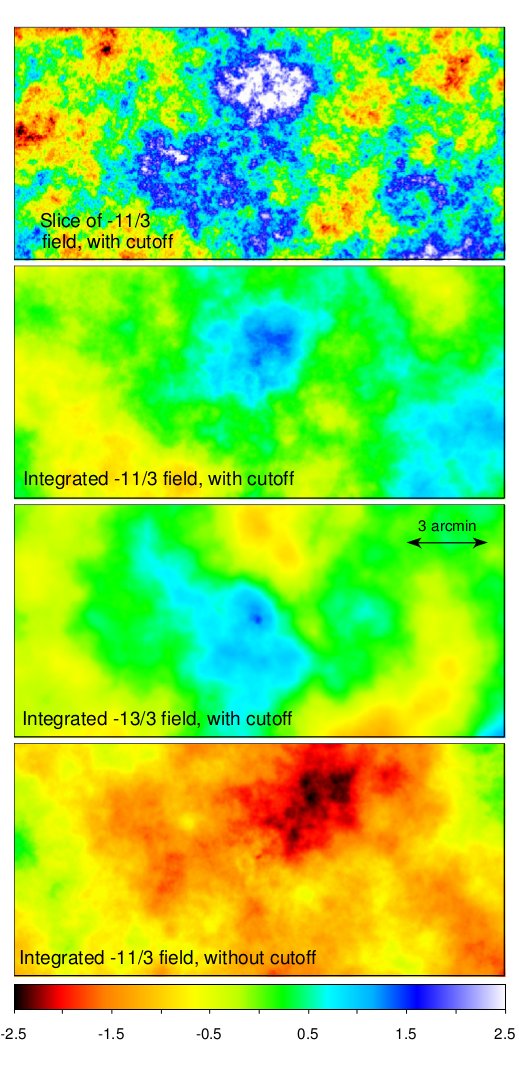}
  \caption{(Top panel) 2D slice through a 3D random field with a power
    spectral index of $-11/3$. We removed power on scales of larger
    than 150~kpc. This field has a standard deviation of 1 in
    3D. (Centre top panel) The same field, after integrating along the
    line of sight, taking account of the cluster emissivity
    profile. (Centre bottom panel) A field with a power spectral index
    of $-13/3$ integrated along the line of sight. (Bottom panel) A
    field with a power spectral index of $-11/3$, with no large scale
    cut-off, integrated along the line of sight.}
  \label{fig:noiseimages}
\end{figure}

The second type of fluctuation model we use is generated using a
powerlaw power spectrum. To make the power spectrum fluctuation
models, we used an inverse fast Fourier transform of a 3D cube where
each pixel was scaled by the power spectrum, but with a normal
distribution of the real and imaginary components.  We used a cube
size of 700 pixels along each axis, which was larger in each dimension
than the \emph{Chandra} image of the cluster.  We took the real
component of the pixels after the inverse transform, scaling them so
that the standard deviation of the fluctuations was of the magnitude
required (the values of such a field have a Gaussian
distribution). Parseval's theorem says that the variance of the pixel
distribution should be equal to area under the power
spectrum. Fig.~\ref{fig:noiseimages} (top panel) shows a slice through
an example 3D field after normalising it to have a standard deviation
of 1.

We examine models with power spectral indices of $-11/3$ and $-13/3$
in 3D. We compare the $-11/3$ models with the
density-squared-sensitive 0.6 to 5~keV band and the $-13/3$
models with the pressure-squared-sensitive 3.5 to 7.5~keV band. We
truncated the power spectrum to zero on scales of larger than 150~kpc.

To compute the projected model we project the 3D field using a
deprojected cluster emissivity model of the galaxy cluster. We take
the average 2D surface brightness profile of the cluster from the
\emph{Chandra} image in the 0.5 to 7~keV band and deproject it
assuming spherical symmetry to compute the emissivity. We integrate
the 3D fluctuation field along a line of sight when it is applied to
the 3D emissivity model. Expressed mathematically, this is
\begin{equation}
  S(x, y) = \int [ 1 + \epsilon F(x, y, z)] E(x, y, z) \: \mathrm{d}z,
\end{equation}
where $S(x,y)$ is the surface brightness on the sky at position $x,y$,
$E(x, y, z)$ is the cluster emissivity model, $F(x, y, z)$ is the
fluctuation field normalised to have a standard deviation of 1 and
$\epsilon$ is the size of the fluctuations required, expressed as a
percentage.  The fluctuations produced by the projection are shown in
Fig.~\ref{fig:noiseimages} (centre top panel). For a particular value
of $\epsilon$, the area under the power spectrum is $\epsilon^2$.

We also show an example field in the centre-bottom panel of
Fig.~\ref{fig:noiseimages} with a powerlaw power spectral index of
$-13/3$, which is more appropriate for the pressure of turbulent
material. This field includes the same large scale cut-off. In bottom
panel we show the effect of removing the large scale cut-off. Large
regions of the image are reduced or enhanced retaining these scales.

If the power spectrum follows a powerlaw,
\begin{equation}
  P(|\mathbf{k}|) \propto |\mathbf{k}|^{-\xi},
  \label{eqn:powspec}
\end{equation}
the $\Delta$-variance should also also give a powerlaw behaviour,
where
\begin{equation}
  \sigma^{2}_\Delta \propto L^{\xi-2},
  \label{eqn:deltavar}
\end{equation}
for length scale $L$, if $0 \le \xi < 6$
\citep{Stutzki98}. Note that this is not realised in
  practice here because the ELFIT cluster model we later subtract from
  the data removes some larger scale power. We use the method with a
Mexican hat filter with a ratio of the sizes of the core and annulus
of 1.5 \citep{Ossenkopf08}. With this kernel, the probed scale is a
factor of 1.075 larger than the size of the kernel. We multiply the
kernel size by this factor when plotting $\Delta$-variance spectra.

\begin{figure}
  \centering
  \includegraphics[width=\columnwidth]{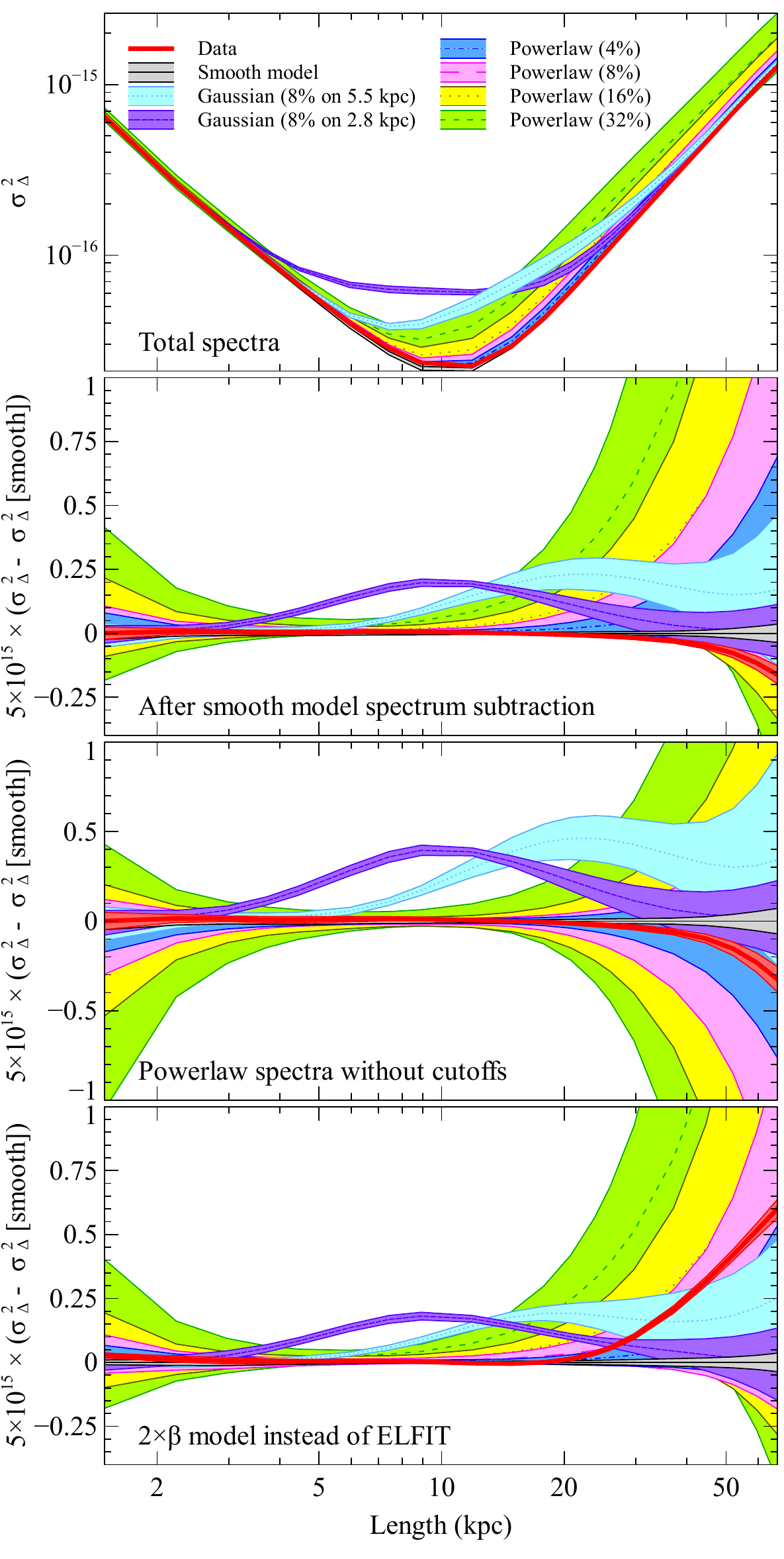}
  \caption{(Top panel) $\Delta$-variance spectra of the 0.6 to 5 keV
    cluster data and ELFIT-based model realisations, between 1 to 6.7
    arcmin radius. The units of the $\Delta$-variance spectra are
    $(\mathrm{photon}\,\mathrm{cm}^{-2}\,\mathrm{s}^{-1})^{2}$. The
    different lines show models with either added Gaussian projected
    fluctuations, or noise with a powerlaw spectral index of $-11/3$.
    (Second panel) Spectra after subtracting the spectrum of the
    smooth model. (Third panel) Comparison with models where we do not
    cut off the power spectrum on long length scales. (Bottom panel)
    Results calculated using a $2\times\beta$ model rather than the
    ellipse model for modelling the cluster surface brightness.}
  \label{fig:spectrum_nosub}
\end{figure}

In the top-panel of Fig.~\ref{fig:spectrum_nosub} we show the
$\Delta$-variance spectrum of the exposure-corrected 0.6 to 5 keV data
in the 1 to 6.7~arcmin radial region. On this and our other
$\Delta$-variance spectra the units are
$(\mathrm{photon}\,\mathrm{cm}^{-2}\,\mathrm{s}^{-1})^{2}$. On short
length scales the signal is dominated by the Poisson noise and on long
scales by the overall morphology of the cluster. In
Fig.~\ref{fig:spectrum_nosub} we also plot the spectra of simulated
cluster images using the 20 ellipse ELFIT model, with different levels
of fluctuations added. For each model we show the mean spectrum for 10
realisations of the model. The shaded regions around each line shows
the standard deviation of the models. The region indicated `smooth
model' is the elliptical model realisation without additional
structure. The Gaussian models add projected fluctuations with a
standard deviation of 8 per cent to the smooth model, on scales of 2.8
and 5.5 kpc. The powerlaw models show the smooth model with $-11/3$
power spectral index fluctuations of the standard deviation shown in
3D. The models show increasing amounts of signal on long
length scales, as the standard deviation is increased, as would be
expected.  The smooth model spectrum is slightly larger than the data
on long length scales. This is more easily seen when subtracting the
smooth model spectrum from the data and other models, seen in the
second panel. Subtracting the smooth model removes most of the effect
of the cluster shape and the Poisson noise.

In the third panel of Fig.~\ref{fig:spectrum_nosub} we show the effect
of removing the 150~kpc cut-off in the model power spectra. With this
cut-off removed, the fluctuations in the models are both positive and
negative. This is because the large scale fluctuations on long length
scales depress or enhance the surface brightness in the region
examined. This negative fluctuation distribution can still be seen in
the models with the cut-off in the second panel. The bottom panel
shows the effect of modelling the cluster with a double-$\beta$ model
fit instead of the ELFIT model. The data show more structure on large
scales than this model (as seen in Fig.~\ref{fig:modelratios}).

\subsection{Spectra after subtraction of smooth cluster model}
\label{sect:subspectra}
We can learn more by removing the dominant contribution on large
scales, by subtracting a smooth cluster model before calculating the
$\Delta$-variance spectra. Some care must be taken when doing this
because the smooth model is based on a fit to the data. We should
treat a smooth model with fluctuations added in the same way as the
data when comparing their spectra. To do this, for each realisation of
a model with fluctuations added, we refit the realised model in the
same way as the data, before subtraction of that fitted model. With
this method, if the fitted model subtracts away real signal from the
data, we will also do it for the simulated model, making comparison of
the spectra valid.

\begin{figure}
  \includegraphics[width=\columnwidth]{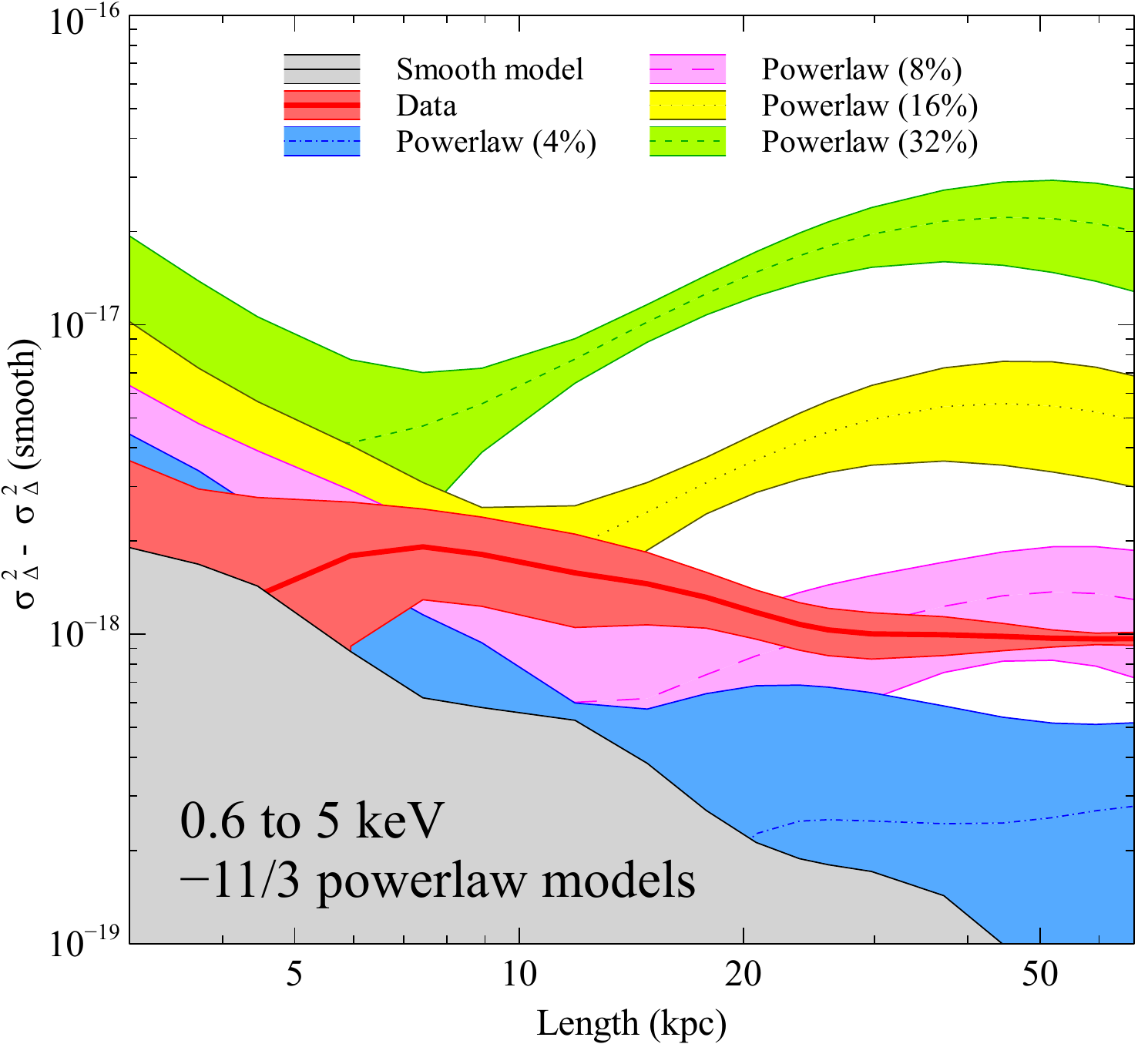}
  \caption{$\Delta$-variance spectra of the data and
    simulated images with increasing levels of powerlaw spectral
    fluctuations added. The plot shows the 0.6 to 5 keV band and 1 to
    6.7 arcmin radial regions.  The smooth ELFIT model was subtracted
    from the images before computation of the spectra. Poisson noise
    in the spectra was removed by subtracting the mean spectrum of
    ELFIT model realisations. The grey shaded region shows the
    standard deviation of the subtracted model. The red line shows the
    spectrum of the data. The blue, pink, yellow and green curves
    show, respectively, the spectra of model clusters with 4, 8, 16
    and 32 per cent, $-11/3$ index power spectra emissivity
    fluctuations added.}
  \label{fig:06_50_powerlaw_spectra}
\end{figure}

\begin{figure}
  \includegraphics[width=\columnwidth]{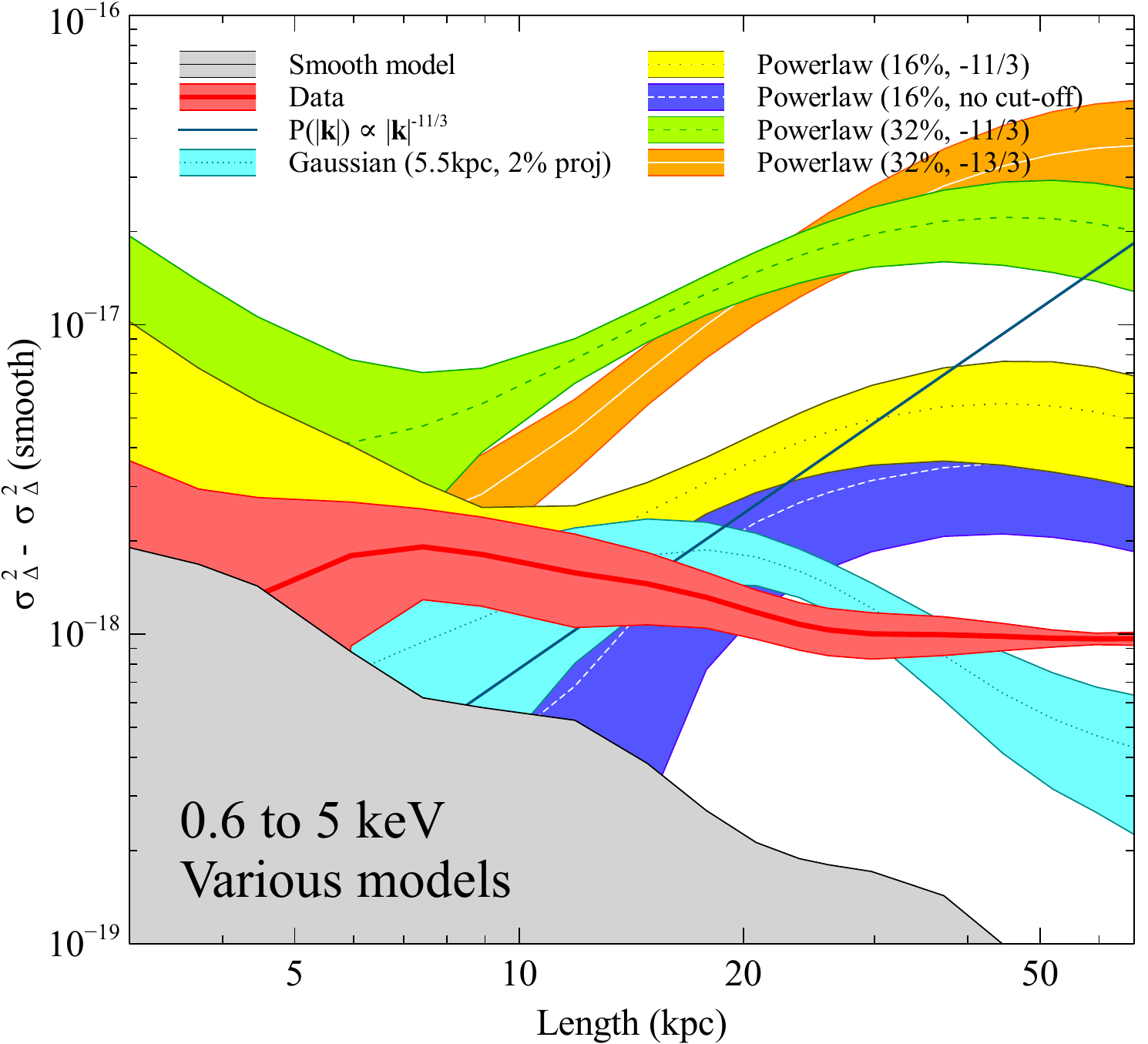}
  \caption{Data spectrum from Fig.~\ref{fig:06_50_powerlaw_spectra},
    plotting additional models. The expected spectrum of $-11/3$
    fluctuations is shown as a solid line. Note that the expected
    spectrum is not achieved for the model spectra because the
    subtraction of the fitted model of the cluster cuts off the
    spectra on long scales. Also shown in cyan is the spectrum of an
    ELFIT model with Gaussian fluctuations at the 2 per cent level on
    scales of 5.5~kpc. The spectra of $-11/3$ powerlaw power spectral
    models with and without the cut-off on large scales, with 16 per
    cent fluctuations, are shown in yellow and purple, respectively.
    The spectra of 32 per cent fluctuation models with $-11/3$ and
    $-13/3$ spectral indicies, are plotted in green and orange,
    respectively. }
  \label{fig:06_50_powerlaw_spectra_xtra}
\end{figure}

  We show the $\Delta$-variance spectra of the data and simulated
  clusters with added powerlaw power spectrum fluctuations in
  Fig.~\ref{fig:06_50_powerlaw_spectra}, examining the 0.6 to 5 keV
  band image between 1 and 6.7 arcmin radius.  After the ELFIT
  subtraction, it is much easier to differentiate the different
  models.  In this graph we plot the spectra logarithmically, as the
  error bounds of the signals are positive. The shaded grey region at
  the bottom of the plot shows the standard deviation of the spectra
  of 10 different ELFIT model realisations. This region is the minimum
  uncertainty on the spectra and shows the region below which signals
  become statistically insignificant. The red line is the spectrum of
  the data and the shaded red region is the standard deviation on the
  smooth model realisation results, used as an estimate of the data
  spectrum uncertainty.

  On this plot we include the spectra for realisations of ELFIT models
  with added powerlaw power spectral fluctuations using a standard
  Kolmogorov index of $-11/3$. We show the model spectra with
  normalisations of 4, 8, 16 and 32 per cent (shown in blue, pink,
  yellow and green, respectively).  The normalisations were defined
  previously to be the standard deviation on the emissivity
  fluctuation distribution.

  It can be see that the fluctuations in the data are detected on
  scales above 5 kpc. The data spectrum does not closely match any of
  the model spectra, but is most similar to the model with 8 per cent
  fluctuations (shown in pink). The shape of the data spectrum is also
  rather flat or even negative. If the data were characterised by a
  power spectral slope, it would be much flatter at around $-1.9$,
  plotting Equation \ref{eqn:deltavar}, rather than $-11/3 \sim -3.7$.

  In Fig.~\ref{fig:06_50_powerlaw_spectra_xtra} we show the data
  spectrum with different spectral models to show the effect of
  various factors.  We show 32 per cent fluctuation models with power
  spectral indices of $-11/3$ (green) and $-13/3$ (orange). Over the
  central ranges of length scale these models can be clearly
  distinguished. The solid line in the plot shows the expected
  spectrum for $-11/3$ fluctuations using
  Equation~\ref{eqn:deltavar}. We do not get the same slope as this
  line on large length scales because the ELFIT models removes some of
  the fluctuations on these scales. We also plot a 16 per cent
  fluctuation model, but show the effect of including (yellow) and
  removing (purple) the cut-off on larger scales in the power
  spectrum. Without the cut-off the distribution the distribution of
  the spectra for the models becomes wider, as discussed in Section
  \ref{sect:deltavar}. By removing the signal above 150~kpc, we boost
  the signal below these length scales when normalised to the same
  standard deviation.  Finally, we show the non-physical Gaussian
  fluctuation model (cyan). It gives a similar signal to the data, but
  has less power on long length scales.

\begin{figure}
  \includegraphics[width=\columnwidth]{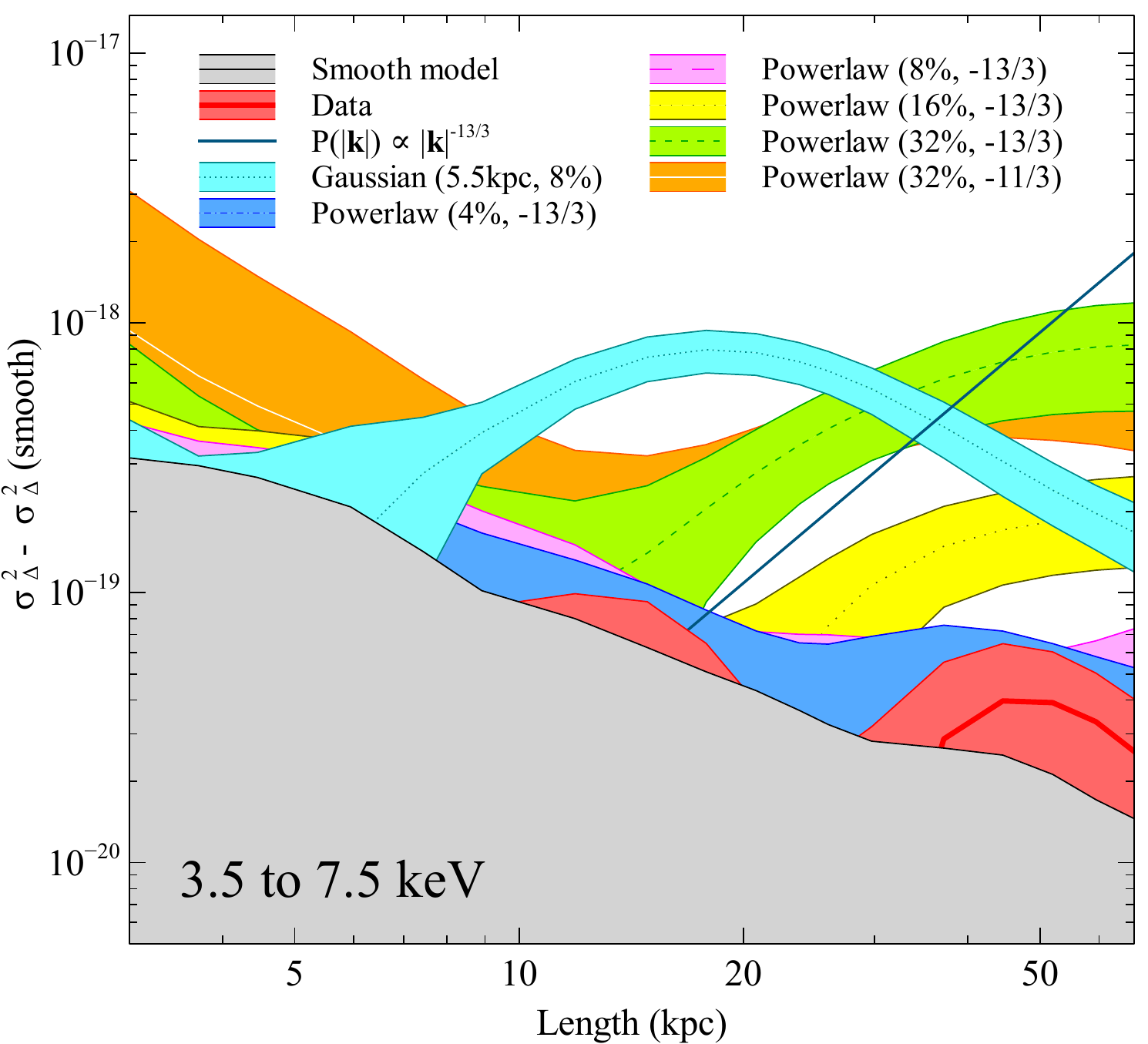}
  \caption{$\Delta$-variance spectra for the 3.5 to 7.5 keV band in
    the 1 to 6.7 arcmin radial region, of the data and model clusters
    after subtracting the ELFIT model.}
  \label{fig:deltaspec_full_35_75}
\end{figure}

If we examine the 3.5 to 7.5 keV data for the same spatial region, we
obtain the spectra in Fig.~\ref{fig:deltaspec_full_35_75}. In this
plot we show the results with powerlaw spectral model fluctuations
using the index of $-13/3$ instead of $-11/3$. The results show little
evidence of a signal above the standard deviation of the smooth model
results, despite the features seen in Fig.~\ref{fig:sim_div_real}. The
results appear consistent with below 8 per cent emissivity
fluctuations however. The spectrum is below the upper bounds of the 4
and 8 per cent models.

\subsection{Spectra for different cluster models}

\begin{figure*}
  \centering
  \includegraphics[width=\textwidth]{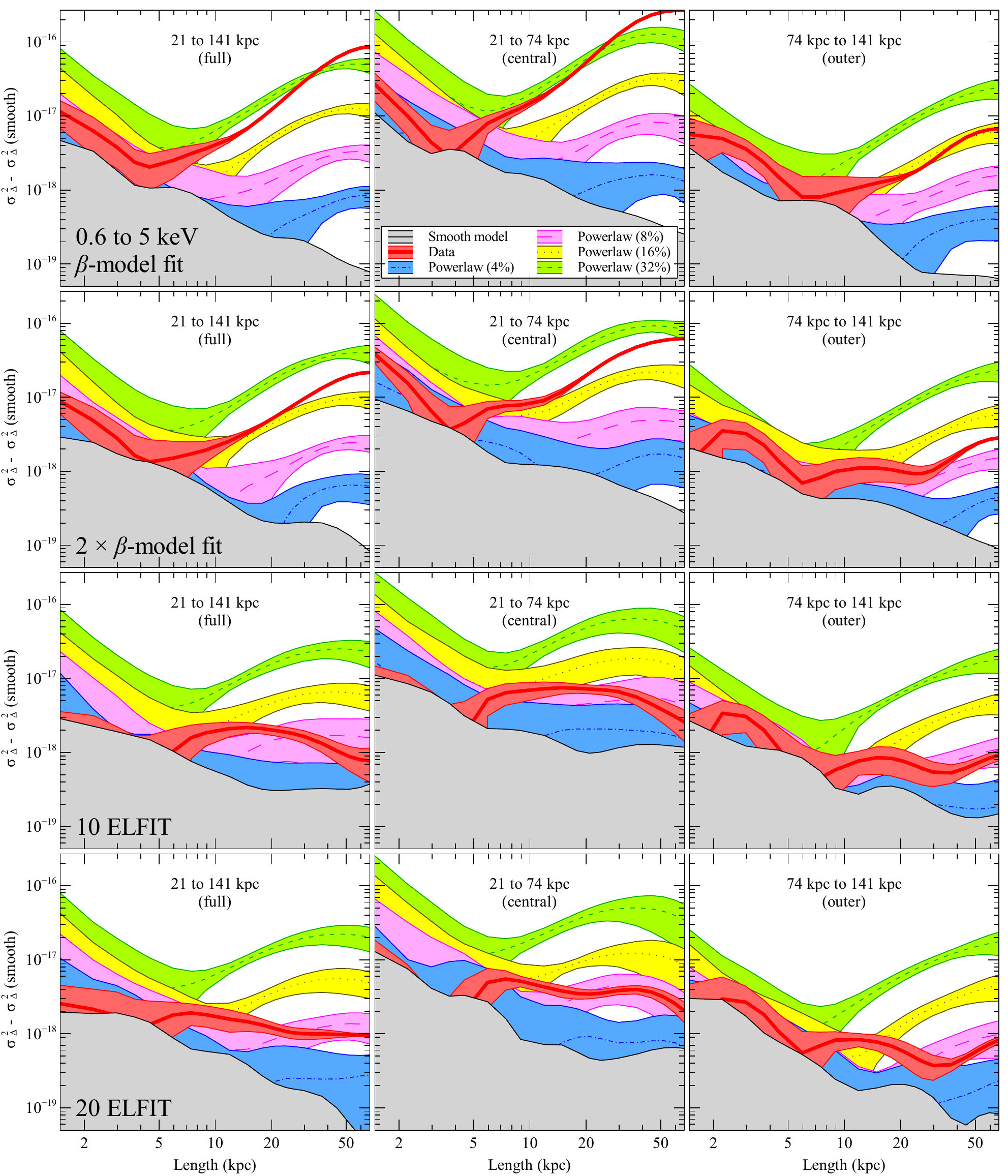}
  \caption{$\Delta$-variance spectra for the 0.6 to 5 keV
    band in the 1 to 6.7 arcmin radial region (left), 1 to 3.5 arcmin
    (centre) and 3.5 to 6.7 arcmin (right) regions, after subtraction
    of the smooth model spectra. The rows show the result of
    subtracting different models for the cluster surface
    brightness. The top row is for a $\beta$ model fit to the surface
    brightness, outside 1 arcmin. The second row is for a 2 component
    $\beta$ model, fitted to the whole \emph{Chandra} region. The
    third and fourth rows show the results using ELFIT models with 10
    and 20 ellipses, respectively.  Within each panel we show the
    spectra of the data (red), a model without any fluctuations
    (grey), models with powerlaw power spectra with an index of
    $-11/3$ (dark blue, pink, yellow and green, with increasing
    normalisation).}
  \label{fig:deltaspec_regions}
\end{figure*}

We can compare how the choice of smooth model for the cluster emission
affects the $\Delta$-variance spectrum. We can also examine whether
the spectra are different between spatial regions or as a function of
radius. In Fig.~\ref{fig:deltaspec_regions} we show the spectra for
the data and simulated data with fluctuations after
subtraction of the single $\beta$ model, a two-component $\beta$ model
and ELFIT models with 10 and 20 ellipses (shown in
Fig.~\ref{fig:contours} and Fig.~\ref{fig:modelratios}). Before
computing the spectra we again also fit the simulated model images
with the model before subtraction of that model to make a fair
comparison between the data and model. The results for different
smooth models are shown by the rows in
Fig.~\ref{fig:deltaspec_regions}. We do the analysis for the standard
1 to 6.7 arcmin radius region (21 to 141 kpc) in the first column, the
central 21 to 74 kpc (1 to 3.5 arcmin) in the central column and the
outer 74 to 141 kpc (3.5 to 6.7 arcmin) region in the final column.

  The choice of smooth model makes a large difference to the resulting
  power spectra normalisation and shape. Using the $\beta$ model fit,
  the data have a $\Delta$-variance spectrum similar to the model with
  the $-11/3$ power spectrum fluctuations, but the normalisation is
  around 32 per cent in the central region and 16 per cent in the
  outer region. With the double-$\beta$ model the normalisations are
  just above 16 and 8 per cent, respectively. With the 10 ELFIT model
  the cluster shows a $\Delta$-variance spectrum which is even lower
  still. The 20 ELFIT model produces normalisations which are around 8
  per cent (see Section \ref{sect:subspectra}), on long scales in both
  the inner and outer regions. On shorter scales there is a peak of
  signal on scales of around 15~kpc in the outer region.

\begin{figure*}
  \centering
  \includegraphics[width=\textwidth]{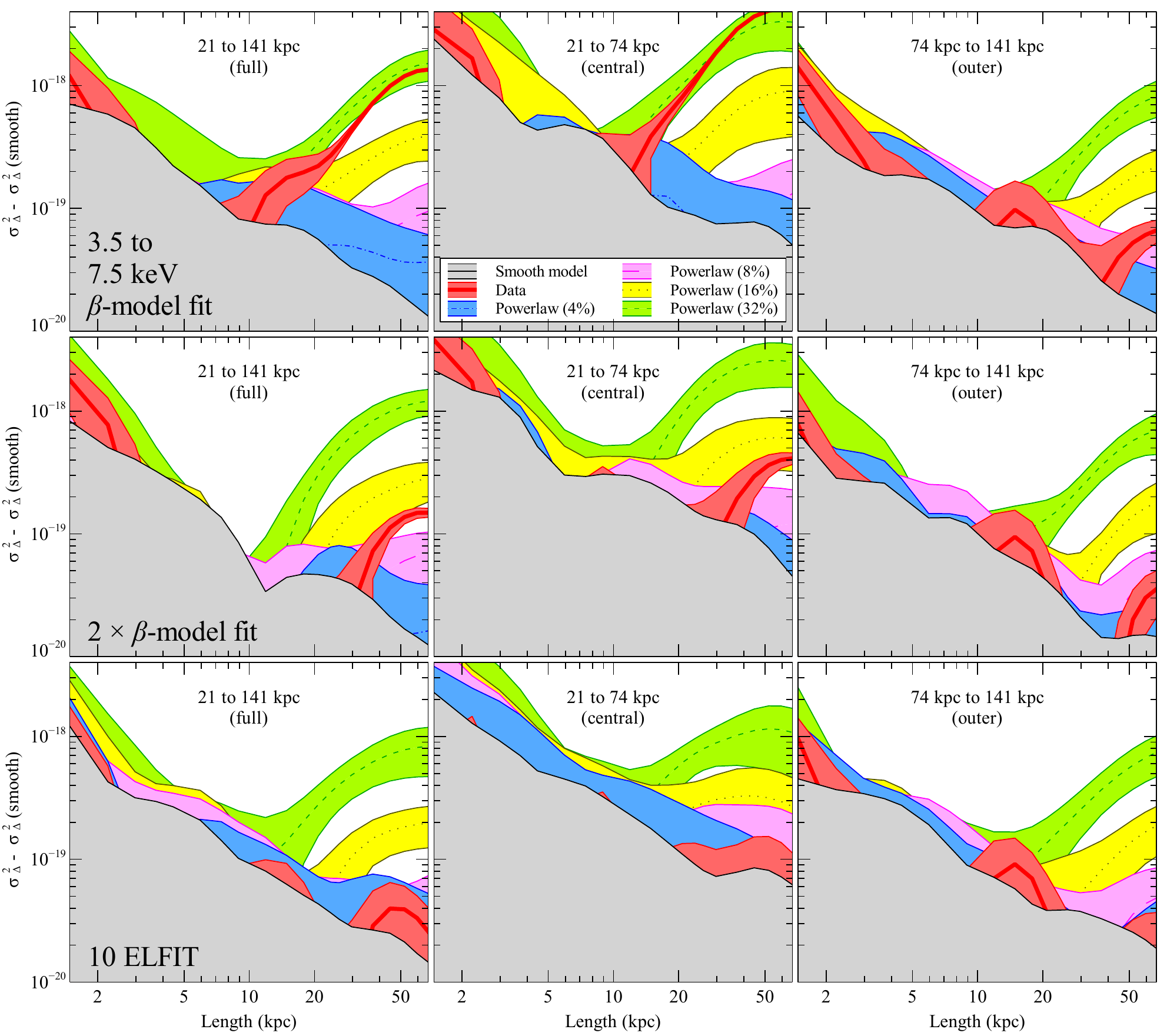}
  \caption{$\Delta$-variance spectra in the 3.5 to 7.5 keV band,
    similarly to Fig.~\ref{fig:deltaspec_regions}. We do not include
    the 20 ELFIT model as there are not enough data. The powerlaw
    power spectrum models use a spectral index of $-13/3$, rather than
    $-11/3$. }
  \label{fig:deltaspec_regions_35_75}
\end{figure*}

We can do a similar analysis in the 3.5 to 7.5 keV band
(Fig.~\ref{fig:deltaspec_regions_35_75}). We use the same type of
modelling as for the 0.6 to 5 keV band, except that the data are too
poor to fit using the 20 ellipse model.  Subtracting a $\beta$ model
fit, the data show similar spectra to those of the $-13/3$ powerlaw
fluctuations with normalisations close to the wider energy band
results, except in the outer region where the normalisation is much
lower. The double-$\beta$ model results show power spectra which are
only significant on large scales, with values between the 8 and 16 per
cent results. Using the ELFIT ellipse fit, there is perhaps a peak in
signal around 15 kpc in the outer part of the cluster, but no evidence
for the peak in the inner part. There is little evidence for any
significant additional fluctuations in any part of the galaxy cluster
in this band using the ellipse model, despite being seen in
Fig.~\ref{fig:sim_div_real}.

\section{Discussion}
\label{sect:discuss}
The surface brightness image of the galaxy cluster appears remarkably
smooth in images (Fig.~\ref{fig:xmm_chandra_compar}). When examining
the full band image, unsharp masking reveals the bright core and an
edge to the north-west (Fig.~\ref{fig:unsharp}). The X-ray emission is
extended in the east-west direction relative to the north-south axis
and the central X-ray peak is offset west from the larger scale
emission (\citealt{Neumann95} and Fig.~\ref{fig:contours}).

  When subtracting the smooth ELFIT model fit from the 0.6 to 5~keV
  data (Fig.~\ref{fig:sim_div_real} top-left panel) there are linear
  residuals which are not present in simulated realisations of the
  smooth model (bottom-left panel).  The magnitude of the residuals is
  around 4 per cent of the surface brightness.  The strongest features
  are linear depressions towards the south-west and south-east
  radiating from the centre. They are also confirmed in
  \emph{XMM-Newton} EPIC-MOS images
  (Fig.~\ref{fig:xmm_chandra_residuals}).  These residuals are also
  present in the harder pressure-sensitive 3.5 to 7.5~keV band
  (Fig.~\ref{fig:sim_div_real} top-right panel).

We show the distribution of surface brightness relative to the ELFIT
smooth model in Fig.~\ref{fig:elldistn0650} and
Fig.~\ref{fig:elldistn3575}. The data show a wider distribution of
pixel value than the ELFIT realisations in all four of the elliptical
regions shown, examined on scales of 4 and 8 pixels (2.8 and 5.6 kpc,
respectively). The data have a narrower distribution than the ELFIT
realisations with added projected Gaussian fluctuations with a
standard deviation of 4 per cent. The data histograms are not
completely consistent with a Gaussian shape, however, and are
typically weighted towards small negative deviations and
counterbalancing large positive deviations.

The measured deviations are dependent on the surface brightness model
subtracted from the cluster image. We examined four different models
in Fig.~\ref{fig:contours} and Fig.~\ref{fig:modelratios} for the 0.6
to 5 keV band. The single and two component elliptical $\beta$ models
are unable to subtract the smooth cluster emission. The models cannot
properly account for the offset of the core, twisting isophotes and
edges seen in the residuals in Fig.~\ref{fig:modelratios}. The
residuals, however, do not look like classical hydrodynamical
turbulence (e.g. Fig.~\ref{fig:noiseimages}). The residuals to the
$\beta$ model fits resemble cold fronts, which are contact
discontinuities, seen in data and cluster simulations
\citep[e.g.][]{Ascasibar06}. These features could be due to residual
sloshing of gas in the potential well. The ELFIT model with 10
ellipses is also unable to properly model the sharp edge immediately
east of the cluster core and further out to the west.

The high level of fluctuations seen in the $\Delta$-variance spectra
of the 0.6 to 5 keV band in the top three rows of
Fig.~\ref{fig:deltaspec_regions} appears mostly due to the large scale
edges and offsets seen in Fig.~\ref{fig:modelratios}. When these
features are subtracted by the 20 ELFIT model the $\Delta$-variance
spectra are significantly reduced on large scales (see
Fig.~\ref{fig:06_50_powerlaw_spectra} or bottom row of
Fig.~\ref{fig:deltaspec_regions}). If the subtraction of
  the ELFIT model removed the turbulence-like powerlaw noise, we would
  be unable to differentiate the models with varying amounts of
  fluctuations in Fig.~\ref{fig:06_50_powerlaw_spectra}.  This is
  because we apply the same model-fitting analysis to the simulated
  data as the real data. The ELFIT modelling removes
contact-discontinuities which would be invisible in a pressure map of
the cluster.

We therefore detect fluctuations in the 0.6 to 5 keV data, but the
magnitude of these features is lower compared to many of our
models. The model with a spectrum nearest to the data on longer scales
has 3D emissivity deviations with an 8 per cent standard deviation and
a power spectrum index of -11/3. The shape of the power spectrum from
the data appears to be flatter, however, with an index of between
$-5/3$ and $-6/3$ at long scales. The amount of power on small scales
is smaller than the lowest powerlaw which has 4 per cent
fluctuations. There is a peak in the $\Delta$-variance spectrum on
scales of around 15 kpc. The shape of the data spectrum is similar to
the Gaussian projected fluctuation model, but is less peaked. We note
that when comparing the spectrum of the data with the model spectra,
that the model spectra are normalised on length scales of zero to
150~kpc.

If we compare the spectra of the inner part of the cluster to the
outer part (centre and right panels of the bottom row of
Fig.~\ref{fig:deltaspec_regions}), then the central region has a
flatter spectrum than the outer region. Ignoring the high small scale
power, the outer region roughly follows the 8 per cent powerlaw power
spectrum model. The central region contributes to the peak in the
total $\Delta$-variance spectrum, peaking on scales of 15 to a few 10s
of kpc.

If we assume that the level of emissivity fluctuations in the 0.6 to 5
keV band has a standard deviation of 8 per cent then the corresponding
standard deviation for density fluctuations is 4 per
cent. This value is surprisingly close to the 5 per cent
  value seen on small scales in in the Coma cluster
  \citep{ChurazovComa11}, despite AWM~7 appearing to be a more relaxed
  object on large scales.

\begin{figure}
  \includegraphics[width=\columnwidth]{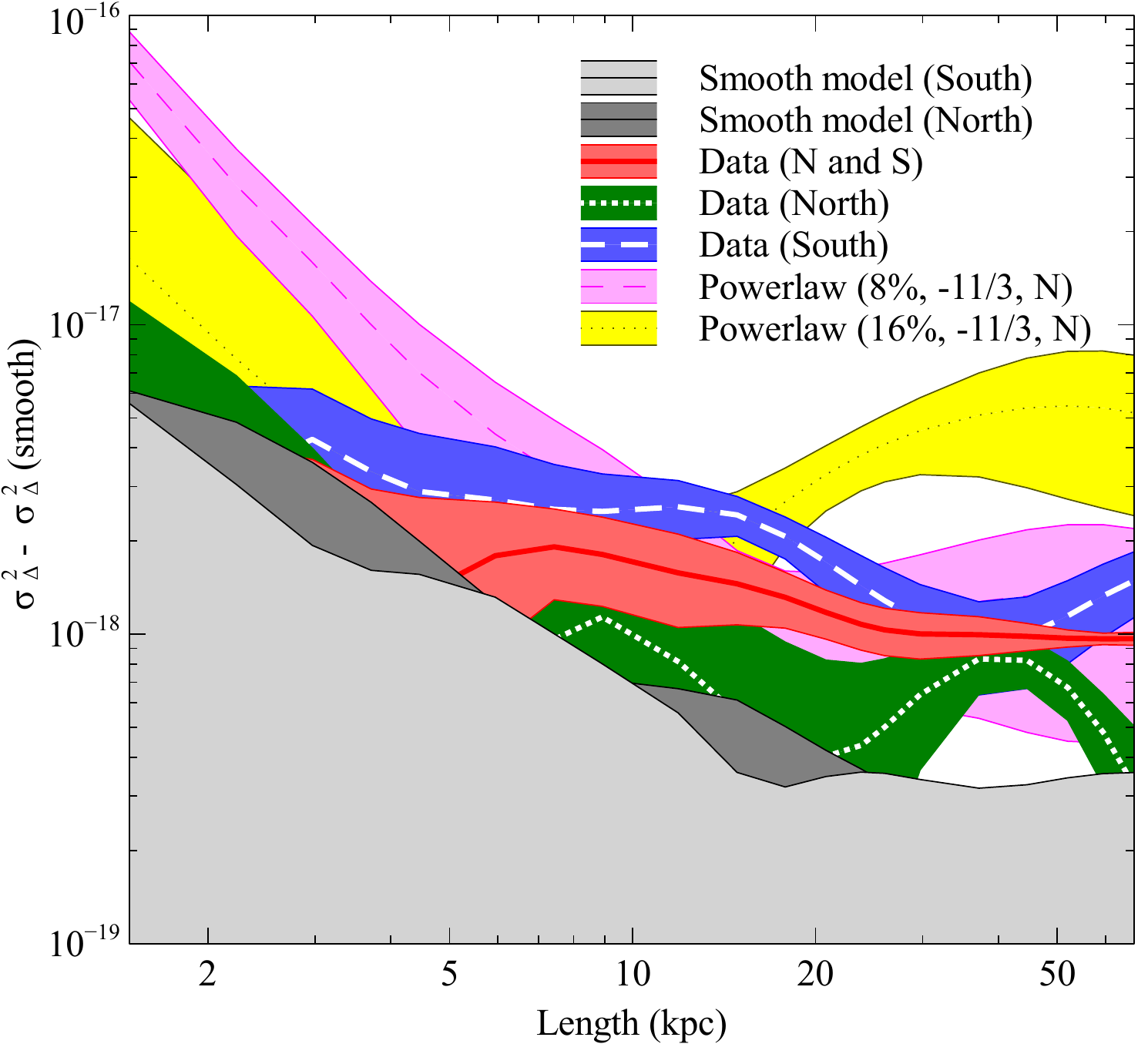}
  \caption{$\Delta$-variance spectra for the northern and southern
    halves of the cluster, examining the radii between 1 and 6.7
    arcmin. An ELFIT model was subtracted before the spectra were
    calculated and the mean smooth model spectrum was subtracted from
    each plotted spectrum, similarly to Fig.~\ref{fig:06_50_powerlaw_spectra}.
    The green (thick dotted) and purple (thick dashed) regions show
    the spectrum of the data in the north and south, respectively. The
    red (thick solid) spectrum is for both halves.  The two grey
    shaded regions at the bottom of the plot are the standard
    deviations of realisations of the ELFIT model in the northern and
    southern halves.The yellow (light dotted) region shows a spectrum
    for a model with powerlaw power spectra with 16 per cent
    fluctuations and an index of -11/3 and the pink (light dashed)
    region is the same with an 8 per cent normalisation.}
  \label{fig:N_S_spectra}
\end{figure}

  We can test whether the radial depressions to the south of the core
  of the cluster are the main contribution to the shape of the
  $\Delta$-variance spectrum. We computed the spectra for the northern
  and southern sectors of the cluster independently in the 0.6 to
  5~keV band, between radii of 1 and 6.7 arcmin, after subtraction of
  the ELFIT model (Fig.~\ref{fig:N_S_spectra}). These spectra have the
  noise removed by the subtraction of the spectrum of a
  model-subtracted realisation of the ELFIT model, as in
  Fig.~\ref{fig:06_50_powerlaw_spectra}. Also shown in the plot is the
  spectrum of the both halves of the cluster, and the spectrum of a
  model with a powerlaw power spectrum fluctuations normalised to have
  16 per cent standard deviation. The southern region, which contains
  the radial depressions, shows more power than in the north on
  virtually all length scales.

  The normalisation of the north spectrum on long scales is roughly
  consistent with the powerlaw power spectrum with a normalisation of
  around 4 per cent. The implied density variations are therefore
  around 2 per cent in the north.

There will be some contribution to surface brightness fluctuations due
to small scale metallicity fluctuations. Such metal features are found
in several other clusters, e.g. Perseus
\citep{SandersPer04,SandersPer07} and Abell 2204
\citep{SandersA220409}. The importance of this effect at the
predominant $\sim 3.5$~keV temperature of this cluster is small. 20
per cent enhancements in metallicity from $0.3\Zsun$ only increase the
0.6 to 5 keV surface brightness by $\sim 2.5$ per cent.

  Examining the pressure-squared-sensitive 3.5 to 7.5 keV band, the
  $\Delta$-variance spectra (Fig.~\ref{fig:deltaspec_full_35_75} and
  Fig.~\ref{fig:deltaspec_regions_35_75}) are harder to differentiate
  from the smooth ELFIT model spectra. On scales of greater than 5 kpc
  they are roughly consistent with powerlaw model emissivity
  variations normalised to 4 or 8 per cent, although there appears to
  be some signal on scales of 15~kpc in the outer region and on longer
  scales in the central region. If the pressure fluctuations are half 
  the emissivity variations, the standard deviation of pressure
  fluctuations is less than 4 per cent.

We therefore find that there are significant fluctuations in the
surface brightness of AWM\,7, but they are of a low level. The
standard deviation of thermal pressure fluctuations is 
  less than 4 per cent. These values are smaller than the lower limit
found for the Coma cluster of 10 per cent \citep{Schuecker04}. AWM\,7,
however, is a much more relaxed cluster than the dynamically disturbed
Coma cluster. AWM\,7 is not a prototypical relaxed, cool-core cluster,
however, as it has a small, cool core and its isophotes are
twisted. We will show in a later paper that the temperature profile is
largely isothermal over the 25 to 250 kpc radius region.

\cite{Churazov10} compared X-ray observations of hot gas and stellar
circular speed for several elliptical galaxies, finding that the
non-thermal pressure, on average, is 20 to 30 per cent of the thermal
gas pressure. This value is much larger than the level of pressure
deviation we could attribute to turbulence. If these elliptical
galaxies had similar levels of turbulence to AWM\,7 then it would
require most of this pressure to be due to other sources such as
cosmic rays or magnetic fields.

Some caution must be exercised when interpreting our results,
however. The conversion from observed surface brightness to density or
pressure depends on the assumption that the dominant pressure
contribution is thermal in origin. Real galaxy clusters are known to
contain relativistic particles and magnetic fields. If these
components of the cluster affect the thermal gas on the scales
examined, the interpretation becomes much less straightforward. In
addition, non-equilibrium thermal processes may become important under
some conditions. If non-thermal effects are important, we would not be
able to simply compare the observed power spectral index against the
Kolmogorov value. It would be very valuable to create mock
observations from simulations incorporating relativistic particles and
magnetic fields, to help interpret our results.

The simulations of \cite{Vazza11} find the ratio of turbulent to
thermal energy in relaxed clusters ranges from 2 to 8 per cent. If we
characterise the detected fluctuations in AWM\,7 as turbulence, our
results are consistent with this range of values. \cite{Vazza11},
however, do not yet probe scales below 25 kpc\,$h^{-1}$, so we
cannot make a direct comparison using the same analysis method as
applied to the data. In addition, the effect of the activity of the
central nucleus is not included in these simulations and may be
important for the shape of spectrum observed.

Some of the signal in the $\Delta$-variance spectrum
  from the central regions may be related to the activity of the
central nucleus. Feedback by the supermassive black holes in galaxy
clusters can create ripples in surface brightness interpreted as sound
waves, seen in the Perseus \citep{FabianPer03,SandersPer07}, Centaurus
\citep{SandersSound08} and Abell 2052 \citep{Blanton09} galaxy
clusters.  Sound waves should give rise to characteristic length scale
in the $\Delta$-variance spectrum at some level, although at their
expected magnitude they are very hard to detect in all but the nearest
and brightest clusters with current telescopes \citep{Graham08}.

In this paper we have not taken advantage of the spectral information
available in the data. We will investigate the distribution of
pressure fluctuations using measured temperature and density in a
future paper. We will also examine the cool X-ray core itself.

\section{Conclusions}
We examine deep \emph{Chandra} and \emph{XMM-Newton} images of the
bright AWM\,7 galaxy cluster. After representing the large scale
surface brightness distribution with a model made by interpolating
ellipses fitted to surface brightness contours, we find residual
features which are approximately radial, have a projected magnitude of
4 per cent in the 0.6 to 5 keV band and are strongest in the south of
the cluster. Examining the 3.5 to 7.5 keV band, which is more
sensitive to pressure variations, we also find evidence for such
features at a similar magnitude.

A histogram of the surface brightness distribution after subtraction
of a smooth model also shows a wider distribution than expected from
the Poisson statistics.

If, instead of the ELFIT model for the surface brightness, we fit a
$\beta$ or double-$\beta$ model to the cluster, the residuals are
considerably larger because of the offset peak in X-ray emission in
this cluster and a number of edges in surface brightness.

  We use $\Delta$-variance spectra to characterise the scales on which
  the fluctuations occur. The magnitude of the fluctuations is roughly
  consistent with a model with powerlaw power spectra with an index
  and a standard deviation of 8 per cent pressure fluctuations in 3D
  emissivity the 0.6 to 5 keV band. The spectrum appears flatter than
  expected, however. The implied density fluctuations have a standard
  deviation of around 4 per cent. In the pressure-squared-sensitive
  3.5 to 7.5 keV band, the 3D pressure fluctuations appear to be 4 per
  cent or less.

  We find that there is a large difference between the signal in the
  $\Delta$-variance spectra between the north and south of the
  cluster. On the longest scales there is roughly a factor of two less
  $\Delta$-variance signal in the north relative to the south.

\section*{Acknowledgements}
The authors thank E.~Churazov for giving a talk making them aware of
the $\Delta$-variance method. ACF acknowledges the support of the
Royal Society.

\bibliographystyle{mnras}
\small
\bibliography{refs}

\end{document}